\documentclass[sigconf, nonacm, pdfa]{acmart}

\usepackage[a-2b]{pdfx}
\usepackage{balance}

\newcommand\vldbdoi{10.14778/3836663.3836676}
\newcommand\vldbpages{3105 - 3118}
\newcommand\vldbvolume{19}
\newcommand\vldbissue{11}
\newcommand\vldbyear{2026}
\newcommand\vldbauthors{\authorsclear}
\newcommand\vldbtitle{\shorttitle} 
\newcommand\vldbavailabilityurl{URL_TO_YOUR_ARTIFACTS}
\newcommand\vldbpagestyle{empty} 

\usepackage{amsmath, amsthm}
    
\usepackage{graphicx, xcolor} 
\usepackage{hyperref, url}
\usepackage{threeparttable}
\usepackage{array}
\usepackage{amsthm}

\usepackage{booktabs, multirow}
\usepackage{caption}
\usepackage{subcaption}
\usepackage{enumitem}
\usepackage{setspace}

\usepackage[most]{tcolorbox}
\tcbuselibrary{theorems}
\usepackage[most]{tcolorbox}
\usepackage{amsmath,amsthm}
\usepackage{xparse} 

\definecolor{rev1color}{rgb}{0, 0.2, 0.6}        
\definecolor{rev2color}{RGB}{147, 28, 117}       
\definecolor{rev3color}{RGB}{239, 135, 51}       
\newcommand{\rOne}[1]{\textcolor{rev1color}{#1}}
\newcommand{\rTwo}[1]{\textcolor{rev2color}{#1}}
\newcommand{\rThree}[1]{\textcolor{rev3color}{#1}}
\renewcommand{\rOne}[1]{#1}
\renewcommand{\rTwo}[1]{#1}
\renewcommand{\rThree}[1]{#1}

\NewDocumentEnvironment{boxedlemma}{O{}}
  {\begin{tcolorbox}[colback=gray!5,
                     colframe=black,
                     boxrule=0.5pt,
                     arc=1mm,
                     top=1mm, bottom=1mm, left=1mm, right=1mm,
                     enhanced]
   \refstepcounter{lem}
   \textbf{Lemma \thelem.} \label{#1}%
   \ignorespaces}
  {\end{tcolorbox}}

\usepackage[ruled, linesnumbered, norelsize]{algorithm2e}

\begin{document}


\title{Error-bounded Point Cloud Compression Using Truncated Octahedron Quantization}

\author{Youyuan Liu$^*$, Longtao Zhang$^\#$, Ruoyu Li$^\#$, Bo Jiang$^*$, Taolue Yang$^*$, Kai Zhao$^\#$, Sheng Di$^\dagger$, Eduard Dragut$^*$, Sian Jin$^*$}
\newcommand{\authorsclear}{Youyuan Liu, Longtao Zhang, Ruoyu Li, Bo Jiang, Taolue Yang, Kai Zhao, Sheng Di, Eduard Dragut, Sian Jin}
\affiliation{%
  \institution{Temple University$^*$, Philadelphia, PA, USA\\
  Florida State University$^\#$, Tallahassee, FL, USA\\
  Argonne National Laboratory$^\dagger$, Lemont, IL, USA
}
}
\email{{youyuan.liu, bo.jiang, taolue.yang, edragut, sian.jin}@temple.edu, {lzhang11, rl13m, kai.zhao}@fsu.edu, sdi1@anl.gov}

\begin{abstract}
With the rapid advancement of large-scale scientific simulations, the massive volume of point cloud data generated has increasingly become a critical bottleneck for scientific storage systems and data management pipelines.
Existing point cloud compression techniques integrated into scientific storage systems are designed for sparse geometry and rely on quantization schemes whose optimality assumptions do not hold for dense data. 
When applied at the compression layer to point clouds, this representation mismatch leads to fundamentally sub-optimal rate-distortion trade-offs that cannot be addressed through parameter tuning or framework-level adaptations.
This mismatch increases storage overhead and limits efficient movement and downstream analysis of simulation outputs.
This issue arises in scientific data management workflows handling large-scale dense particle datasets. 
State-of-the-art compression methods fail to fully exploit the redundancies inherent in such data.

We address this limitation by developing a theory of point cloud compressibility for dense data, characterizing fundamental rate-distortion behavior at the representation layer. Guided by this analysis, we introduce \texttt{XnYZip}, an error-bounded lossy compressor based on provably optimal Truncated Octahedron quantization, combined with a locality-aware encoding pipeline using space-filling curves and run-length encoding. Experiments on large-scale scientific datasets demonstrate consistent storage and throughput improvements, achieving up to $3\times$ higher compression ratios, $2.2\times$ faster compression, and $1.2\times$ faster decompression compared to state-of-the-art point cloud compressors under same distortion.


\end{abstract}

\maketitle

\pagestyle{\vldbpagestyle}
\begingroup\small\noindent\raggedright\textbf{PVLDB Reference Format:}\\
\vldbauthors. \vldbtitle. PVLDB, \vldbvolume(\vldbissue): \vldbpages, \vldbyear.\\
\href{https://doi.org/\vldbdoi}{doi:\vldbdoi}
\endgroup
\begingroup
\renewcommand\thefootnote{}\footnote{\noindent
This work is licensed under the Creative Commons BY-NC-ND 4.0 International License. Visit \url{https://creativecommons.org/licenses/by-nc-nd/4.0/} to view a copy of this license. For any use beyond those covered by this license, obtain permission by emailing \href{mailto:info@vldb.org}{info@vldb.org}. Copyright is held by the owner/author(s). Publication rights licensed to the VLDB Endowment. \\
\raggedright Proceedings of the VLDB Endowment, Vol. \vldbvolume, No. \vldbissue\ %
ISSN 2150-8097. \\
\href{https://doi.org/\vldbdoi}{doi:\vldbdoi} \\
}\addtocounter{footnote}{-1}\endgroup

\ifdefempty{\vldbavailabilityurl}{}{
\begingroup\small\noindent\raggedright\textbf{PVLDB Artifact Availability:}\\
The source code, data, and/or other artifacts have been made available at \url{https://github.com/Karshilov/XnYZip}.
\endgroup
}



\keywords{Data Compression, Data Management, Data Quality, Point Cloud}

\section{Introduction}
\label{sec:introduction}

Data compression has become a fundamental component of modern database and data management systems~\cite{Cheng2014,Gray2005,hdf5,Kersten2011,hu2025dcc,hu2025icde}, where it is widely used to reduce storage footprints~\cite{compInDB,oracle}, accelerate queries \cite{Fenget18,Fenget21,ZhiyuanGK01}, and mitigate I/O bottlenecks in data-intensive workloads~\cite{Jiao2022}. While built-in lossless compression techniques~\cite{compInDB,oracle} ensure exact data recovery, they provide limited reduction factors for floating-point scientific datasets. General-purpose lossy compressors and time-series codecs, such as those designed for natural media (e.g., JPEG~\cite{jpeg1992short,wallace1992jpeg,taubman2012jpeg2000}) or specialized data stores like SummaryStore~\cite{summarystore}, often lack quantifiable error guarantees, making them unsuitable for scientific workloads that require controlled fidelity and predictable distortion~\cite{wang2022tac}. As a result, error-bounded lossy compression has emerged as a critical storage-layer mechanism in scientific data management systems~\cite{hdf5,hdf5-filter,hdf5filter-sz,modelarDB,xia2024,xia2026timevaryingvectorfieldcompression}, enabling substantial space savings while preserving analysis-ready representations.

This challenge is amplified by high-performance computing simulations in domains such as molecular dynamics and cosmology, which increasingly generate massive point cloud datasets at unprecedented scale~\cite{HACC,nyx,cesm-atm,EXAALT2021}. As exascale applications leverage modern supercomputing platforms to run large ensembles of simulations, they routinely produce dense collections of particles stored at high precision, pushing the memory, storage, and I/O capacities of existing data infrastructures to their limits. For example, the EXAALT project’s ParSplice framework executes large numbers of molecular dynamics replicas, each producing thousands to millions of particles per snapshot, resulting in data volumes that quickly reach terabyte to petabyte scale over the course of a single simulation campaign~\cite{EXAALT2021}. The sheer magnitude of such point cloud data makes storage, transfer, and post-analysis prohibitively expensive without effective compression support.

Despite the increasing adoption of error-bounded lossy compression in data management systems, existing approaches do not perform uniformly well across different classes of point cloud data. Point clouds generated by scientific simulations are typically much denser than those arising in computer graphics or autonomous driving, where data tends to be relatively sparse. Many current point cloud compressors, including state-of-the-art solutions~\cite{google_draco,Schwarz2019GPCC}, are primarily designed with such sparse data in mind. Beyond density, these two categories of point clouds also differ in their structural regularity: scan-based point clouds often exhibit a minimum sampling resolution and more regular spatial patterns, while scientific point cloud data may exhibit either regular or irregular structures depending on the underlying physical model and simulation setup. In contrast, point clouds produced by molecular dynamics and similar simulations are often less constrained by fixed sampling steps and, therefore, display weaker spatial regularity. Thus, compression techniques optimized for sparse and more regular point clouds are often ineffective when applied to dense simulation data with diverse structural characteristics.
More fundamentally, \textit{this limitation reflects a mismatch between how compression fidelity is evaluated and how quantization is performed}. The performance of point cloud compressors is predominantly assessed using L2-based metrics such as Mean Squared Error (MSE)~\cite{Schwarz2019GPCC,mpeg_pcc_tmc13,MPEG_I_PCC,2017fan}, yet their underlying quantization schemes almost universally operate by quantizing each coordinate axis independently. 
This scalar quantization strategy is geometrically equivalent to partitioning space with a cubic lattice, which is provably suboptimal for minimizing MSE~\cite{voronoi}. Consequently, \textit{the quantization strategies employed by existing compressors are not  optimized, from a geometric perspective, for the very fidelity metrics by which they are judged.}

To address this fundamental mismatch, we develop a theory for the compressibility of dense point cloud data in storage systems.
Many existing compressors share a common architecture: distortion is introduced through quantization (or an equivalent mechanism), followed by encoding schemes that exploit correlations across coordinate axes. 
However, there is currently no unified theory that links these design choices to data compressibility under L2-based fidelity metrics. 
Our theory fills this gap by deriving compressibility bounds under L2-based MSE, explicitly relating lattice cell geometry to the expected distortion, and yielding density-aware predictions that align with empirical measurements. 
Guided by—and directly instantiating—our theory, we propose \texttt{XnYZip}, an error-bounded lossy compressor for point cloud data.
We replace the traditional cube quantization with \textbf{Truncated Octahedron (TO) quantization}, which our theory identifies as optimal for minimizing MSE under vector quantization. 
\texttt{XnYZip} employs a Space-Filling Curve (SFC)~\cite{Hilbert1891,morton1966} to serialize 3D TO indices into a 1D sequence, preserving spatial locality to enable effective predictive and entropy coding. In addition, an adaptive run-length encoding module is incorporated to exploit high-duplication scenarios common at low bitrates. The main contributions of our work are outlined as follows:
\begin{itemize}[left=0em, itemsep=0pt, topsep=2pt]
    \item We develop a theory of dense point cloud compressibility under L2-based fidelity metrics, relating quantization geometry to distortion and rate-distortion behavior.
    \item Guided by this theory, we design \texttt{XnYZip}, an error-bounded lossy point cloud compressor based on TO quantization and locality-aware encoding.
    \rOne{\item We propose an MPI-based distributed compression framework for \texttt{XnYZip}, enabling parallel compression at scale without sacrificing compression ratio and demonstrating favorable iso-efficiency.}
    \item We provide theoretical analyses for the major components of \texttt{XnYZip}, including closed-form derivations and optimality results.
    \item We evaluate \texttt{XnYZip} on multiple large-scale scientific datasets and demonstrate consistent improvements over state-of-the-art compressors, achieving up to $3\times$ higher compression ratios, $2.2\times$ faster compression, and $1.2\times$ faster decompression under the same distortion levels.
 
\end{itemize}

\section{Background and Related Work}
\label{sec:background}

\subsection{Error-Bounded Lossy Compression}

Lossy compression provides high compression ratios by accepting controlled information loss, in contrast to lossless methods that preserve data exactly. A critical distinction in this domain is between compressors optimized for perceptual quality, such as JPEG~\cite{wallace1992jpeg}, and those designed for numerical or geometric data that provide \textbf{strict, user-defined error bounds}~\cite{sz3,sz16,sz17,sz18,zfp}. The latter guarantees that the reconstruction error will not exceed a specified threshold,
ensures data fidelity in many scientific applications.

Data compression is a core technique in database systems for managing large-scale scientific datasets. Error-bounded lossy compression has emerged as a practical compromise between storage efficiency and analytical fidelity, enabling substantial data reduction while enforcing explicit reconstruction error bounds. Widely used compressors such as SZ~\cite{sz3} and ZFP~\cite{zfp} are effective for structured numerical arrays, while MGARD~\cite{ainsworth2018multilevel} leverages multiresolution decomposition to support hierarchical error guarantees. In practice, such compressors are commonly deployed as storage-layer primitives in scientific data management pipelines, including database-backed analytics and file-based systems such as HDF5~\cite{hdf5}.

However, the effectiveness of these lossy compression frameworks are highly dependent on the structure of the data. Unordered and sparse data such as 3D point clouds lack the regular connectivity that these compressors rely on, requiring specialized algorithms.

\begin{figure}[]
\centering
\begin{subfigure}{0.3\linewidth}
\includegraphics[width=0.65\textwidth]{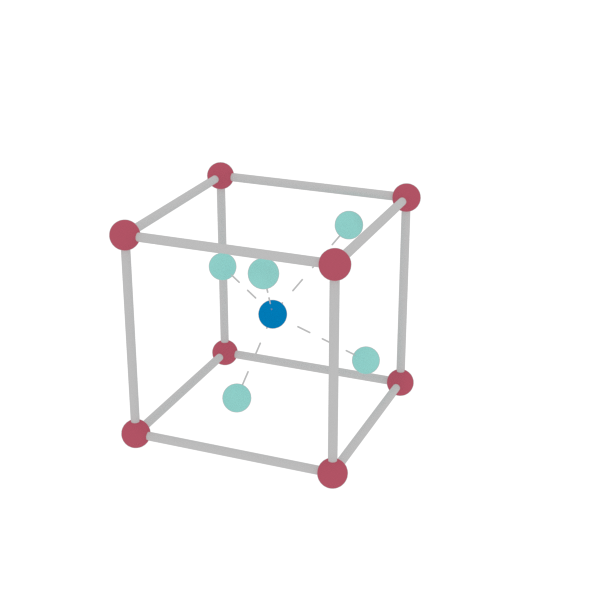}
\subcaption{Cube}
\label{fig:cube-q}
\vspace{1em}
\end{subfigure}
\begin{subfigure}{0.3\linewidth}
\includegraphics[width=0.65\textwidth]{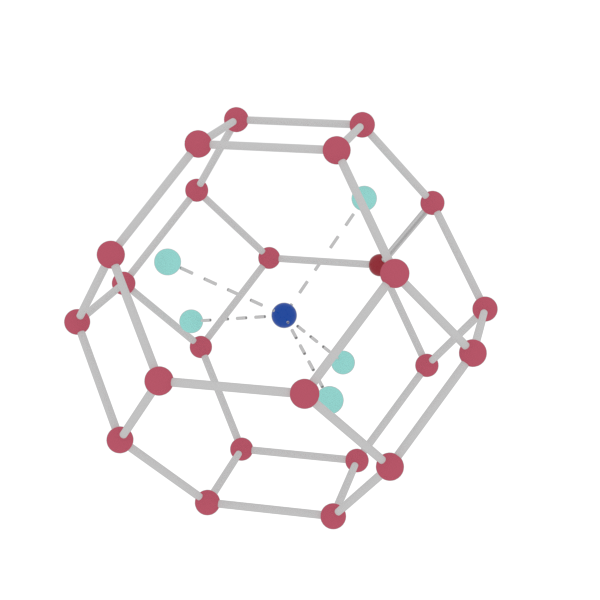}
\subcaption{Truncated Octahedron}
\label{fig:to-q}
\end{subfigure}
\begin{subfigure}{0.3\linewidth}
\includegraphics[width=0.65\textwidth]{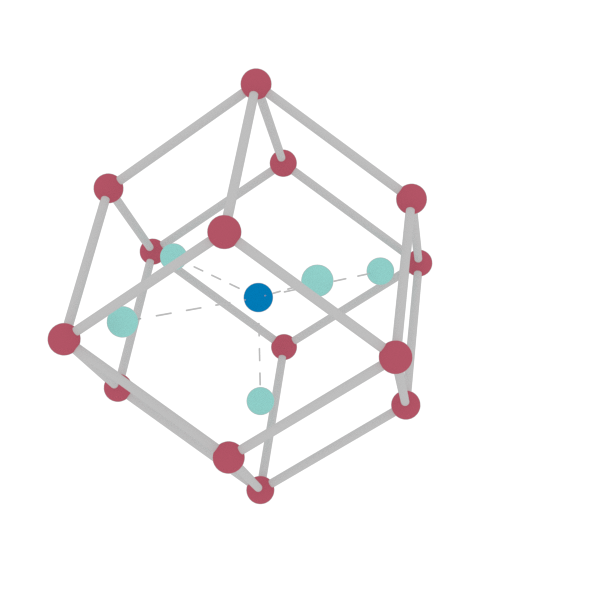}
\subcaption{Rhombic Dodecahedron}
\label{fig:rd-q}
\end{subfigure}
\begin{subfigure}{0.3\linewidth}
\includegraphics[width=0.65\textwidth]{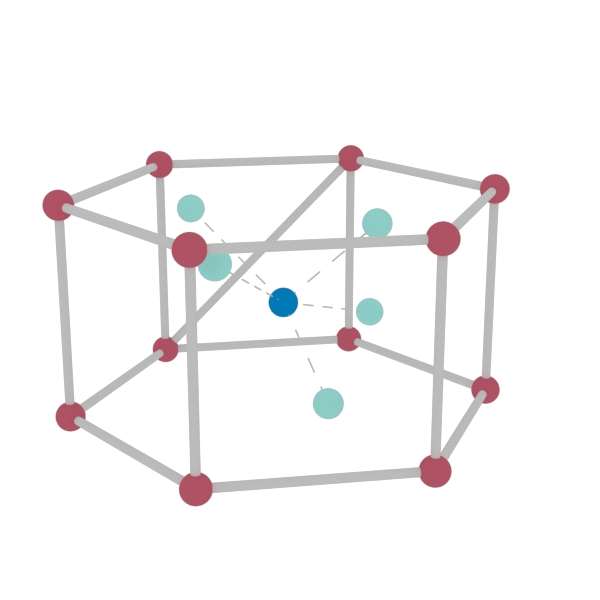}
\subcaption{Hexagonal Prism}
\vspace{1em}
\label{fig:hp-q}
\end{subfigure}
\begin{subfigure}{0.3\linewidth}
\includegraphics[width=0.65\textwidth]{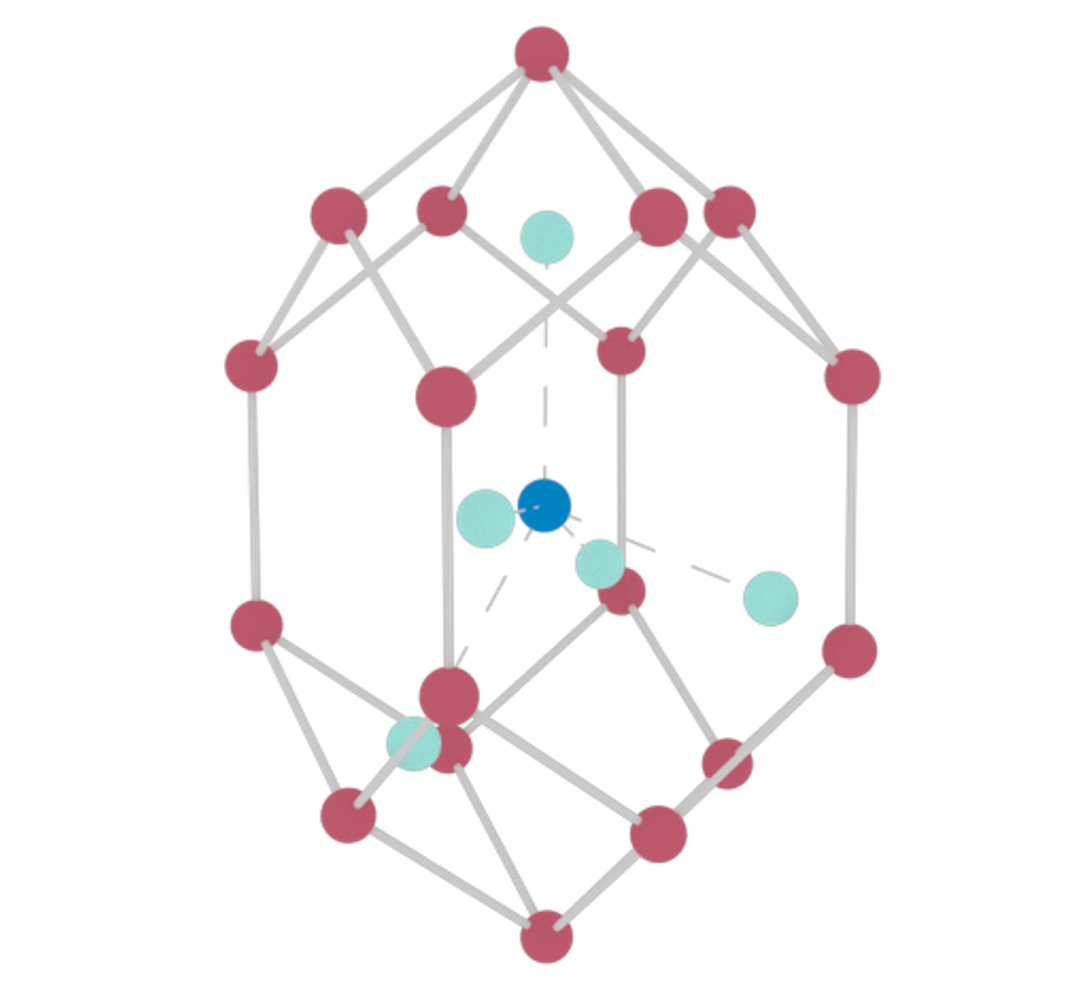}
\subcaption{Elongated Dodecahedron}
\label{fig:ed-q}
\end{subfigure}
\caption{
A visual comparison of quantization schemes using the five convex polyhedra that can tile 3D space by translation.
}
\label{fig:quant-schemes-viz}
\end{figure}

\subsection{Point Cloud Compression}

A 3D point cloud is a set of points in 3D space with geometric and attribute data, whose lack of explicit connectivity poses a fundamental compression challenge~\cite{cao2019survey}. The need for compression is driven by massive data volumes, as large-scale dynamic scenes can easily generate hundreds of millions to billions of points per second in raw acquisition~\cite{cao2019survey}.

Existing approaches can be broadly categorized into two main paradigms based on their data organization. The first paradigm, \textbf{projection-based compression}~\cite{Ochotta2004,Daribo2012,Lien2010,Merkle2007}, maps 3D data into 2D planes to leverage highly optimized video codecs. While exemplified by MPEG's V-PCC~\cite{VPCC} and capable of high compression ratios, these methods inherently rely on orthogonal projections that partition space into square pixels, often introducing resampling artifacts unsuitable for high-precision scientific analysis. The second paradigm, \textbf{direct 3D compression}~\cite{Huang2006,Kammerl2012,Kathariya2018,Kitago2006,Gumhold2005,Taubin1998}, operates directly in 3D space. This broad category includes methods using tree structures, such as octrees in MPEG G-PCC, or explicit quantization grids to organize spatial data, employing specialized traversals to exploit locality.

For scientific particle and point cloud data, LCP~\cite{LCP} represents the state-of-the-art error-bounded lossy compressor tailored to unordered point sets, employing multi-level reordering strategies to exploit spatial locality. While effective in practice, LCP and related approaches rely primarily on heuristic design choices and do not provide a principled link between quantization geometry and rate--distortion optimality. Existing work on data compressibility has largely focused on general-purpose compression or structured scientific arrays, with limited attention to dense volumetric point clouds or particle data~\cite{jin2022improving}. Hence, current systems offer little theoretical guidance on how quantization geometry influences compression efficiency and distortion behavior in point cloud settings.

Despite the algorithmic differences between these paradigms, 
virtually all error-bounded lossy point cloud compressors share a fundamental \textbf{geometric consensus}: the use of the \textbf{cubic lattice} for space partitioning. Whether through recursive octant subdivision or explicit Cartesian quantization, the rate--distortion performance of existing methods is implicitly bounded by the geometric efficiency of the cube. This suggests that the bottleneck lies not only in prediction algorithms, but also in the quantization geometry itself.

\rOne{The quality of the reconstructed point cloud $\hat{P}$ against the original point cloud $P$ is assessed using several key metrics. \textbf{Peak Signal-to-Noise Ratio (PSNR)}, derived from the MSE, is a classic metric for measuring distortion, but it requires a direct point-to-point correspondence. It is defined as:
$$
\text{PSNR} = 10 \cdot \log_{10}\left(\frac{Diag^2}{\frac{1}{N}\sum_{i=1}^{N}|p_i - \hat{p}_i|^2_2}\right),
$$
where $Diag$ is the diagonal length of the original point cloud's bounding box. PSNR is the standard fidelity metric used throughout our evaluation, as the L2-based MSE that it summarizes is the distortion that our error-bounded compressor directly controls.}

\subsection{Lattice Quantization and Space-Filling}

Quantization is the foundational process in lossy compression, which \textbf{partitions} the continuous space $\mathbb{R}^n$ into a set of disjoint cells ${R_i}$ that completely cover the space, i.e., a \textbf{tessellation}. Quantization maps all points within a given cell to a single representative value. A particularly principled approach is \textbf{lattice quantization}, which maps any given point in $\mathbb{R}^n$ to the nearest node in a lattice, a regular repeating grid of points.

Formally, a lattice $L$ is a discrete set of points generated by all integer linear combinations of a set of basis vectors $\{\mathbf{v}_1, \dots, \mathbf{v}_n\}$:
$$
L = \left\{\sum_{i=1}^{n} k_i \mathbf{v}_i \mid k_i \in \mathbb{Z}\right\}.
$$
The lattice results in different space-filling partitions, such as a tessellation by cubes or by Truncated Octahedra (TO), shown in Figure~\ref{fig:quant-schemes-viz}.
The Voronoi region of a lattice node contains points closer to that node than any other and forms a space-filling polytope.

The problem of determining the most efficient lattice in terms of rate--distortion performance is a classic subject in information theory~\cite{hires,voronoi}. Building upon this theoretical foundation, we formalize and extend the analysis to explicitly prove the optimality conditions relevant to our quantization framework, thereby bridging classical lattice theory with practical compressor design.


After lattice quantization, a continuous point cloud is transformed into a sparse set of occupied integer voxel coordinates. To encode these coordinates efficiently, the 3D locations are serialized into a 1D sequence before predictive and entropy coding~\cite{Michael1995}. A space-filling curve (SFC) provides such a 3D-to-1D ordering while preserving spatial locality~\cite{Hilbert1891}: points that are close in 3D space are more likely to remain close in the serialized sequence. Better locality leads to smaller differences between consecutive indexes and therefore a lower-entropy sequence.

\begin{figure}[]
\centering
\begin{subfigure}{0.3\linewidth}
\includegraphics[width=0.75\textwidth]{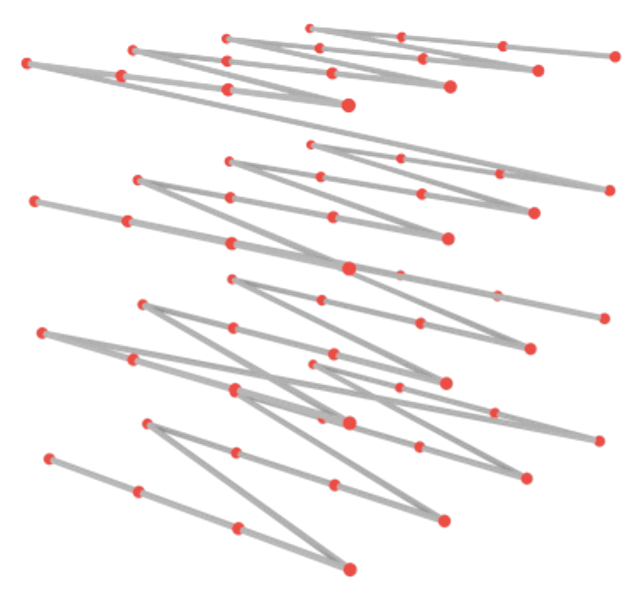}
\subcaption{Row Major Ordering}
\label{fig:row-major}
\end{subfigure}
\begin{subfigure}{0.3\linewidth}
\includegraphics[width=0.75\textwidth]{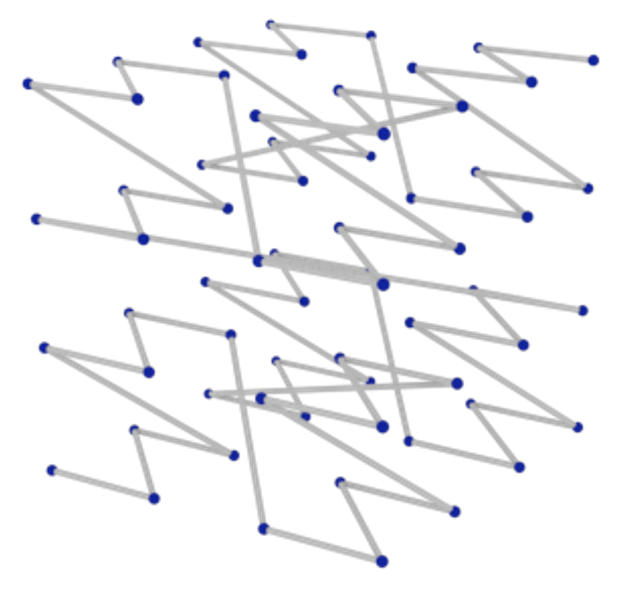}
\subcaption{Z-order Curve}
\label{fig:z-order}
\end{subfigure}
\begin{subfigure}{0.3\linewidth}
\includegraphics[width=0.75\textwidth]{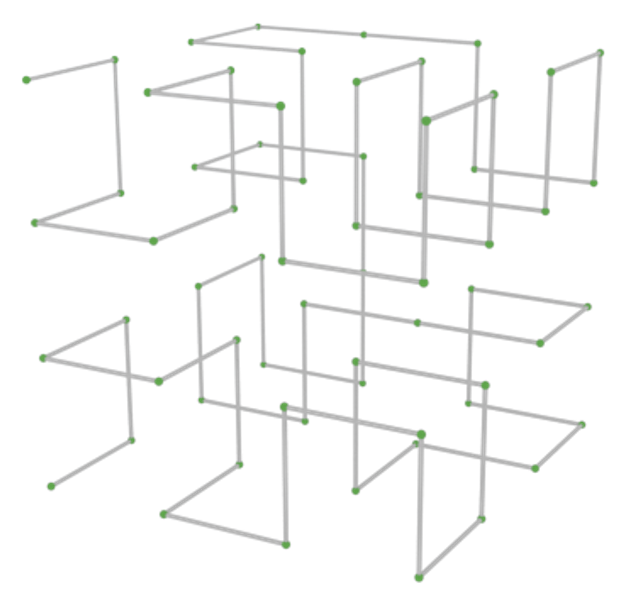}
\subcaption{Hilbert Curve}
\label{fig:hilbert}
\end{subfigure}
\caption{
Visual comparison of three traversal paths on a $4 \times 4 \times 4$ grid. Row-major ordering has frequent long jumps, Z-order reduces these jumps, and Hilbert ordering provides stronger spatial locality.
}
\label{fig:sfc}
\end{figure}

Figure~\ref{fig:sfc} compares three representative traversal orders. Row-major ordering is simple but creates large jumps at row and plane boundaries, which hurts compression. Z-ordering~\cite{morton1966,Tropf1981MultimensionalRS}, improves locality by interleaving coordinate bits, but still contains discontinuities across regions. Hilbert ordering usually provides stronger locality than Z-order~\cite{Hilbert1891}, reducing large jumps in the serialized sequence. Therefore, we use Hilbert ordering to generate a more compressible first-order difference sequence after quantization.

\subsection{Other Related Compression Methods}

\rOne{Several application-aware error-bounded compressors have also been proposed. QPET~\cite{Jiao2022,qpet} preserves user-specified Quantities of Interest (QoIs), while TspSZ~\cite{tspsz} preserves the topological skeleton of vector fields, including critical points and separatrices.
These methods address a complementary problem to ours: they focus on \emph{what} downstream features should be preserved, while our work focuses on \emph{how} quantization geometry affects rate--distortion behavior in dense point cloud data.}

\rThree{Another line of work targets floating-point columns in data management systems. ALP~\cite{alp} provides high-throughput lossless compression for tabular floating-point columns, and BUFF~\cite{buff} supports SIMD-friendly queries under user-specified bit precision. A recent survey~\cite{fpsurvey} studies this regime in detail. 
Although effective for one-dimensional numeric streams, these methods are less effective for 3D point clouds because they encode axes independently and do not exploit spatial structure.
}

Vector compression, such as Product Quantization (PQ)~\cite{jegou2011pq} and RabitQ~\cite{rabitq}, are mainly designed for high-dimensional embeddings and approximate nearest-neighbor search. They prioritize retrieval accuracy rather than reconstruction fidelity, and their coarse quantization is not suitable for scientific and industrial point clouds that require strict geometric error bounds. Therefore, our work complements these methods by providing a quantization-geometry view of error-bounded point cloud compression.

\begin{table}[]
    \centering
    \footnotesize
    \caption{List of Symbols and Abbreviations}
    \label{tab:nomenclature}
    \begin{tabular}{c l || c l}
    \hline\hline
    \textbf{Symbol} & \textbf{Meaning} & \textbf{Symbol} & \textbf{Meaning} \\
    \hline
    
    $P$       & Point Set            & $G(\cdot)$  & Dimensionless second moment \\
    $V$ & Volume     & $N$     & Number of points \\
    $Enc(\cdot)$    & Encoder   & $Dec(\cdot)$ & Decoder \\
    d  & Distance      & $Diag$ & Diagonal length of point cloud \\
    $\Lambda$ & Voxel Set & $p$ & A single point \\
         $\lambda$     & A single voxel  &    $\mathcal{E}$     &  Total square error        \\      
    $\mathcal{P}$ & A polyhedron & $M$ & Number of voxel \\
         $S$     & Symbol  &    $H$     & Entropy        \\ 
         $I$     & Mutual information  &    $c$     & Centroid point        \\ 
    $E(\cdot)$     & Error  &    $\mathcal{R}$     & Partition/Region       \\ 
    $Q$ & Quantizer & $D(\cdot,\cdot)$ & Distortion \\
    \hline\hline
    \multicolumn{2}{c}{\textbf{Abbreviation}} & \multicolumn{2}{l}{\textbf{Explanation}} \\
    \hline
    \multicolumn{2}{c}{TO}  & \multicolumn{2}{l}{Truncated Octahedron} \\
    \multicolumn{2}{c}{MSE}  & \multicolumn{2}{l}{Mean Squared Error} \\
    \multicolumn{2}{c}{RLE}  & \multicolumn{2}{l}{Run-length Encoding} \\
    \multicolumn{2}{c}{SFC}  & \multicolumn{2}{l}{Space-Filling Curve} \\
    \multicolumn{2}{c}{PSNR} & \multicolumn{2}{l}{Peak Signal-to-Noise Ratio} \\
    \hline\hline
    \end{tabular}
\end{table}

\section{Problem Formulation}
\label{sec:problemformulation}

We formulate point cloud compression as a 3D quantization problem and use this formulation to motivate our choice of the BCC lattice and the resulting TO quantization.

\subsubsection*{Problem Definition}
The objective of our project is to develop a lossy compressor for point cloud data that is \textbf{natively optimized for the geometric fidelity metrics} commonly used in the field. The fundamental problem is to find an encoder $Enc$ and a decoder $Dec$ that minimize the rate $R$ (in bits per point) for a given target distortion $D_{target}$. \rTwo{We focus on the fixed-distortion setting because scientific post-analysis typically requires a user-defined error guarantee, whereas traditional media compression often fixes the bitrate.}
The input is a set of 3D points, $P = {\mathbf{p}_1, \dots, \mathbf{p}_N}$ where $\mathbf{p}*i \in \mathbb{R}^3$.
The encoder $Enc$ maps $P$ to a compressed bitstream of length $|Enc(P)|$, and the decoder $Dec$ recovers the \emph{reconstructed} point cloud $P' = Dec(Enc(P))$. The rate $R$ is a function of $|Enc(P)|$ and the number of points $N$, while the distortion $D(P, P')$ measures the dissimilarity between $P$ and its reconstruction $P'$; \rTwo{the error bound is enforced on this reconstruction, not on the bitstream, which is why $P'$ must appear in the formulation.}
The optimization problem is formulated as:
\begin{equation}
\min R = \frac{|Enc(P)|}{N} \quad \text{s.t.} \quad D(P, P') \le D*{target}
\end{equation}

\subsubsection*{Quantization as the Loss Model}
Transform-based methods, such as those derived from the Discrete Fourier Transform (DFT) or Discrete Cosine Transform (DCT), have proven effective for error-bounded compression on structured data like images or regular grids~\cite{zfp,cuZFP,jpeg1992short}.
In contrast, a general point cloud is an \textit{unordered set of coordinates}, where point indices are arbitrary and lack physical meaning.
Consequently, applying a DFT/DCT to such an index-based sequence yields a spectrum that is not an invariant property of the underlying geometry.

Quantization is the cornerstone of error-bounded lossy compression for point clouds.
It partitions the continuous space $\mathbb{R}^3$ into a finite set of disjoint voxels ${V_i}$ and maps all points within a voxel to a single representative point $\mathbf{c}_i$.
Crucially, this process inherently provides an error-bounding mechanism: the maximum possible error for any point is geometrically constrained by the dimensions of its corresponding voxel, e.g., its covering radius.
\textit{The problem of designing an optimal error-bounded compressor for 3D point cloud is therefore transformed to finding the optimal quantization lattice.}

\subsubsection*{Linking Geometry, Distortion, and Rate}

The choice of quantization lattice governs the trade-off between distortion $D$ and rate $R$.

\paragraph{Formalizing the Quantized Representation}
After quantization, the original point cloud $P$ is transformed into a multiset of $N$ voxel identifiers, $\Lambda = {\lambda_1, \lambda_2, \dots, \lambda_N}$. This multiset is not an Independent and Identically Distributed (i.i.d.) source; it carries significant \rOne{\textbf{spatial redundancy}: the geometric coherence of the underlying point cloud introduces strong statistical dependence among neighboring voxel identifiers.}

\paragraph{Formalizing Rate via a Two-Part Information Model}
The theoretical minimum rate $R$ is determined by the true entropy of this structured source. We model the total information content using two components: \textbf{structural information}, which specifies \textit{which} $M$ of the $N_{\text{total}}$ possible voxel locations are occupied, and \textbf{statistical information}, which describes the frequency distribution of the $N$ points among these $M$ occupied locations.

Under a fixed distortion constraint, a smaller number of occupied voxels $M$ reduces the number of possible spatial configurations and thus lowers the structural complexity. It also typically leads to a more skewed frequency distribution for the $N$ points, which lowers the statistical information.
Assuming a locality-preserving encoding that approaches the entropy bound, minimizing the number of occupied voxels $M$ yields a lower achievable rate $R$:
$\rTwo{\min(M) \implies \min(R)}$

\paragraph{The Optimal Quantization Problem}
For a given target distortion $D_{\text{target}}$, a more efficient quantization geometry will require a smaller number of voxels, $M$, to represent the point cloud. Based on the derivation above, a smaller $M$ results in a representation with lower total information content and thus a lower theoretical compressed bitrate.
Therefore, \textit{the problem of finding the optimal quantizer is formalized as finding the lattice that minimizes the number of occupied voxels $M$ for a given target distortion.}

\subsubsection*{The Encoding Problem: From Geometry to Information}

The quantization stage solves the geometric representation problem, transforming the continuous point cloud into a discrete set of occupied voxel locations. The subsequent problem is to losslessly encode this set by exploiting its inherent redundancies. This requires addressing two types of redundancy: \textbf{statistical redundancy}, which comes from the non-uniform frequency distribution of the symbols, and \rOne{\textbf{spatial redundancy}}, which comes from the statistically dependent correlation of voxel locations in space.

\paragraph{Quantifying Spatial Information}
We define the \textbf{Spatial Information} within a sequence of voxel locations $S$ as the degree to which it deviates from being i.i.d. This can be quantified by the \textbf{Mutual Information} between adjacent elements in the sequence:
\begin{equation}
I(S_i; S_{i-1}) = H(S_i) - H(S_i | S_{i-1})
\end{equation}
where $H(S_i)$ is the zeroth-order entropy and $H(S_i | S_{i-1})$ is the first-order conditional entropy. A high mutual information indicates strong adjacent-symbol dependence and predictability, while a value near zero indicates that the sequence is approximately i.i.d.
In our encoding pipeline, locality-preserving ordering and delta coding reduce this dependence and convert much of the spatial redundancy into a lower-entropy residual stream, which can then be compressed by entropy coding.

\subsection{Quantifying the Compressibility}
\label{sec:quant_comp}
While the TO lattice offers a theoretical geometric advantage, realizing this gain in practice depends on the data distribution relative to the quantization scale. To characterize the regime where our method outperforms traditional approaches, we introduce two analysis metrics: \rOne{\textit{Point Density}} and \textit{Distribution Deviation}.

\noindent \textbf{(1) Point Density ($\rho$).}
This metric measures the average occupancy of the quantization cells. For a given error bound $\epsilon$, let $M_{occ}$ be the number of non-empty voxels. The density is defined as:
\begin{equation}
\rho(\epsilon) = \frac{N}{M_{occ}(\epsilon)}
\end{equation}
The geometric packing advantage of the TO lattice is realized primarily in the \textbf{Dense Regime} ($\rho \gg 1$), where multiple points cluster into single voxels. If $\rho \to 1$, quantization degrades to coordinate enumeration, and the voxel geometry becomes less significant.

\noindent \textbf{(2) Distribution Deviation ($\delta$).}
The theoretical optimality of lattice quantization relies on a \textbf{Spatial Uniformity Assumption}, where points are assumed to be uniformly distributed within a voxel, as commonly used in high-resolution quantization theory~\cite{voronoi,hires}. Under this assumption, the $L_2$ error follows a theoretical distribution $Q_{sim}$ dictated by the voxel geometry.

To quantify the validity of this assumption on real-world datasets, we measure the divergence between the \textit{empirical} $L_2$ error distribution ($P_{emp}$) and the \textit{theoretical} distribution ($Q_{sim}$) using the \textbf{Kullback-Leibler (KL) Divergence}:
\begin{equation}
\delta(\epsilon) = D_{KL}(P_{emp} || Q_{sim}) = \sum_{k=1}^{K} p_k \log \frac{p_k}{q_k}
\end{equation}
$\delta \approx 0$ indicates good agreement with the uniformity assumption. A large $\delta$ indicates strong structural bias or boundary effects that violate spatial uniformity, reducing the geometric advantage.

\section{Methodology}
\label{sec:Methodology}

\begin{figure}[]
    \centering
    \includegraphics[width=0.75\linewidth]{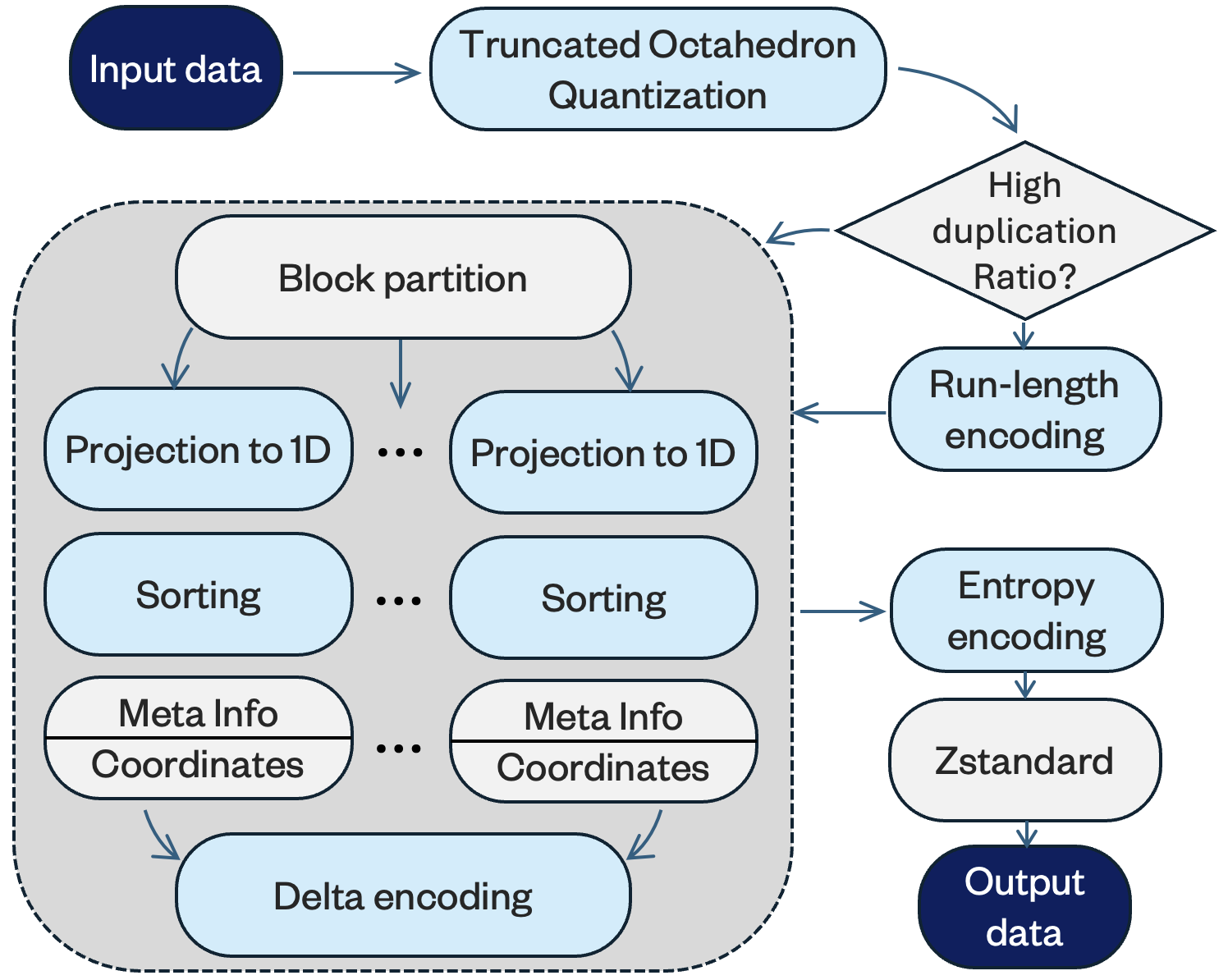}
    \caption{The overview of \texttt{XnYZip}.}
    \label{fig:overview}
\end{figure}

In this section, we introduce our proposed framework, \texttt{XnYZip}, shown in Fig.~\ref{fig:overview}. We then establish the theoretical foundation for our choice of quantization geometry through a series of lemmas.

\subsection{Optimal Quantization Geometry}
\label{sec:optimal-quant-geom}

\subsubsection{Assumption}
\begin{figure}[]
    \centering
    \begin{subfigure}{0.48\linewidth}
        \includegraphics[width=\textwidth]{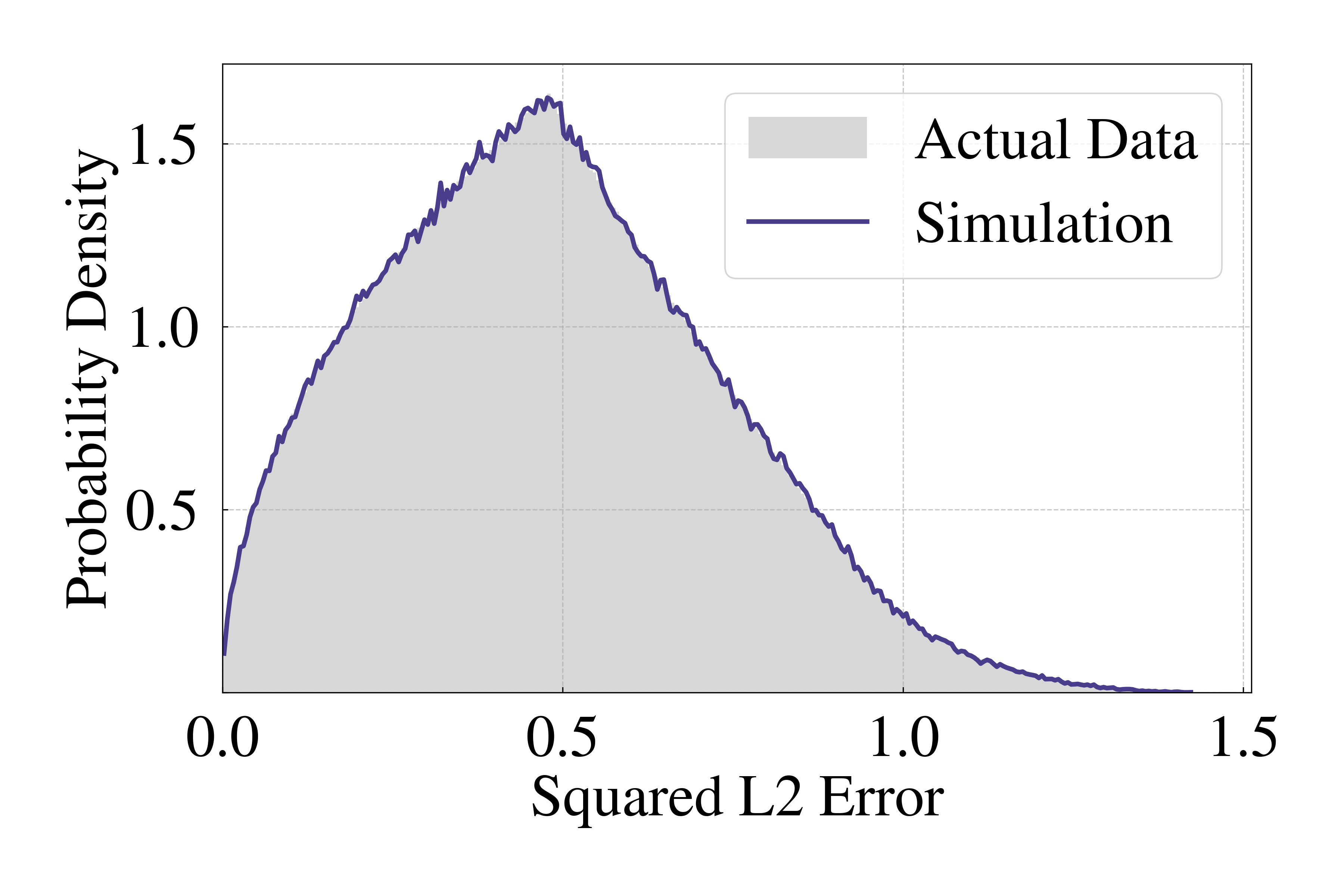}
        \subcaption{Cube quant, USGS dataset.}
        \label{fig:usgs-cube}
    \end{subfigure}
    \begin{subfigure}{0.48\linewidth}
        \includegraphics[width=\textwidth]{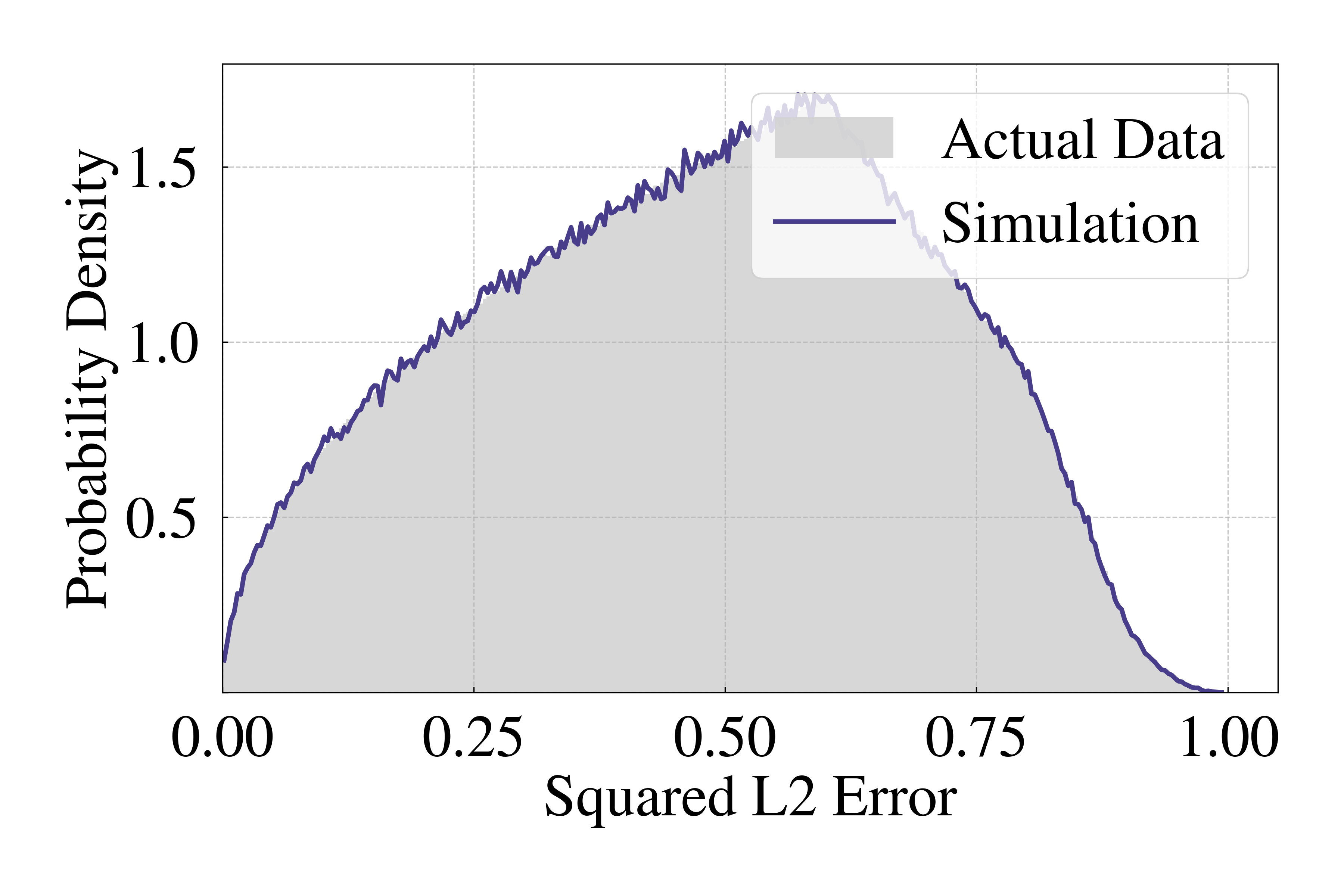}
        \subcaption{TO quant, USGS dataset.}
        \label{fig:usgs-to}
    \end{subfigure}
    \begin{subfigure}{0.48\linewidth}
        \includegraphics[width=\textwidth]{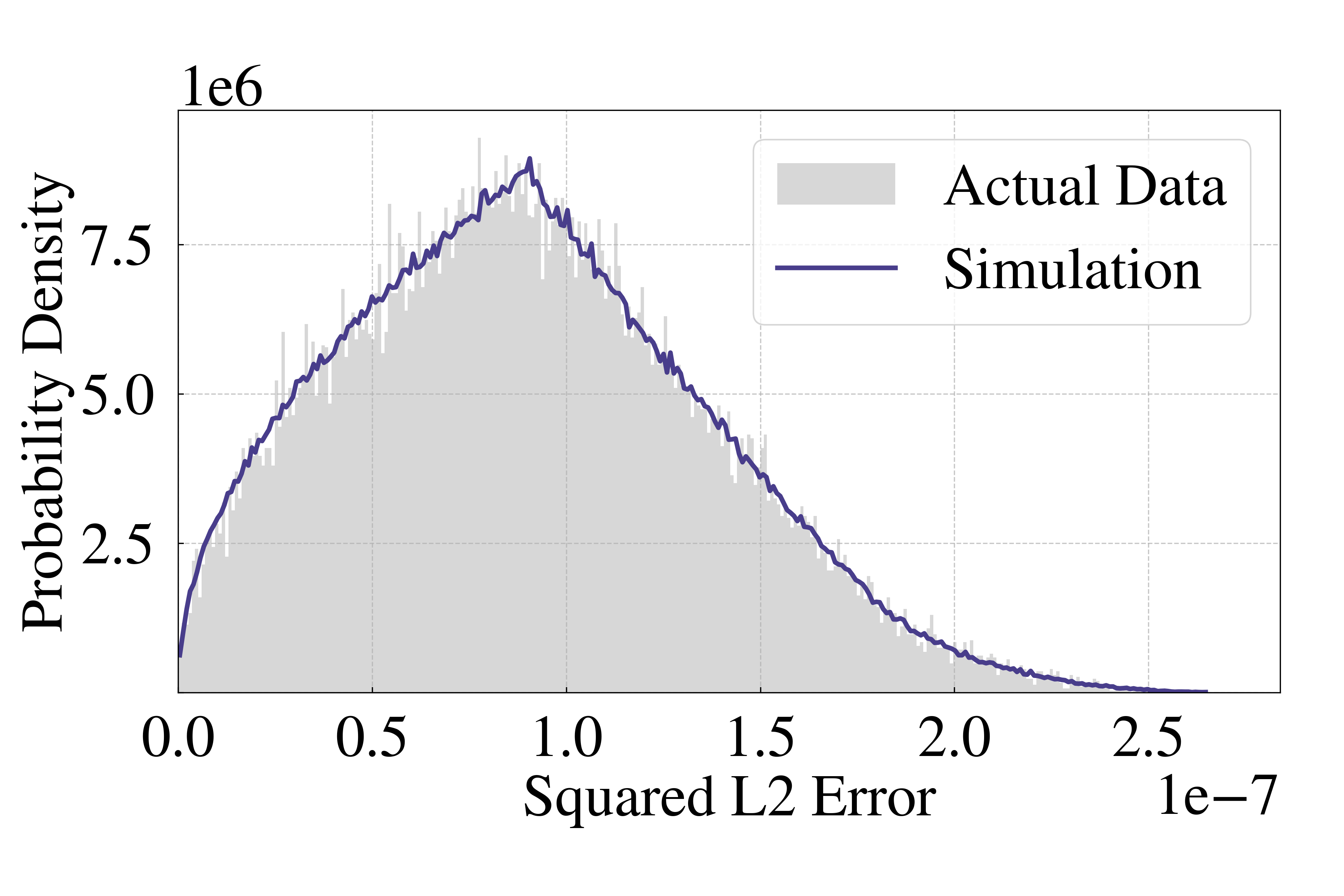}
        \subcaption{Cube quant, Stanford Bunny.}
        \label{fig:bun-cube}
    \end{subfigure}
    \begin{subfigure}{0.48\linewidth}
        \includegraphics[width=\textwidth]{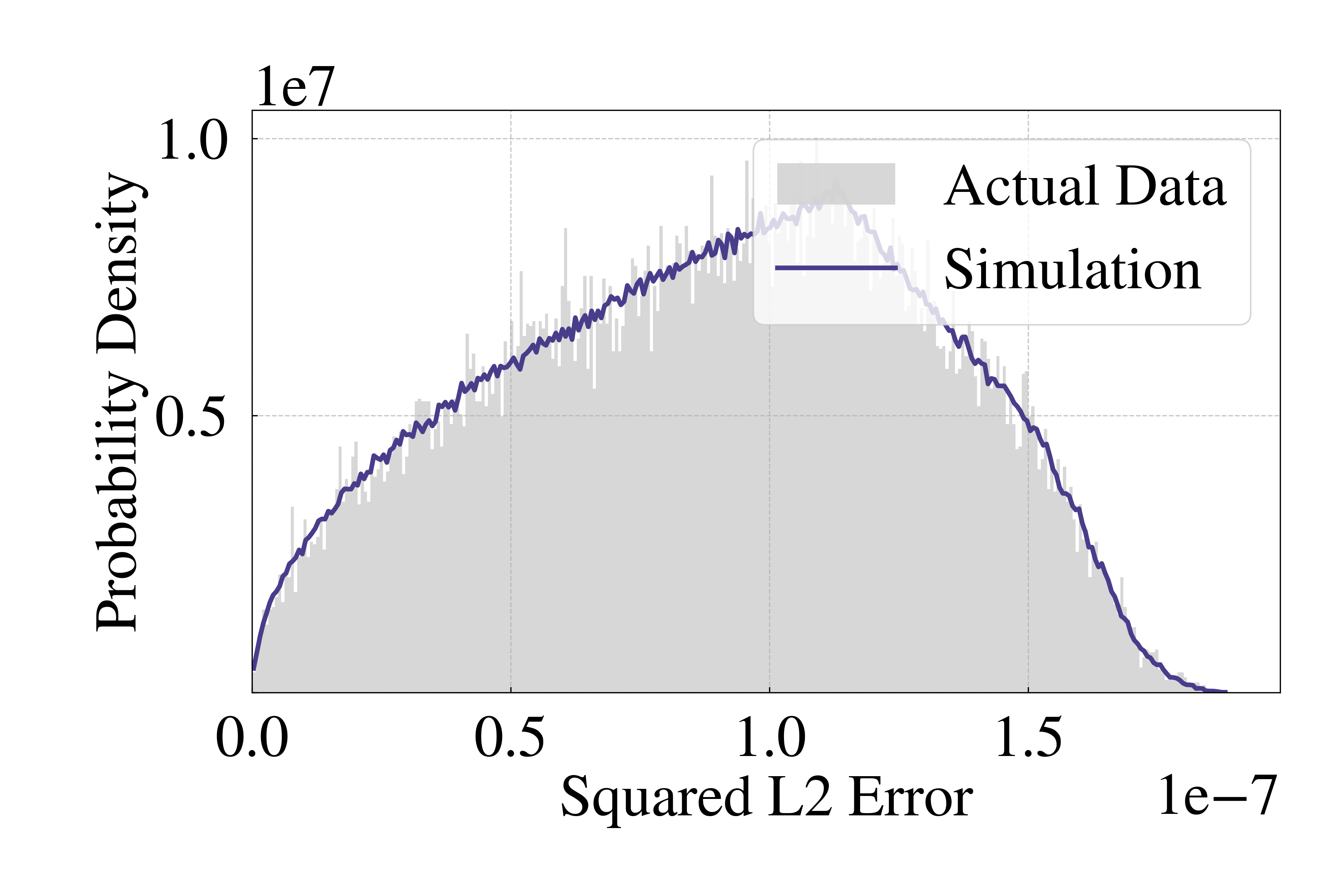}
        \subcaption{TO quant, Stanford Bunny.}
        \label{fig:bun-to}
    \end{subfigure}
    \caption{
    Comparison of the actual quantization error distributions (histograms) against the theoretical error distributions (Solid blue line).
    The comparison is shown for both the Cube and Truncated Octahedron (TO) quantizers on the dense USGS dataset (top row) and the sparse Stanford Bunny dataset (bottom row), with reasonable $eb$.
}
    \label{fig:error-distributions}
\end{figure}

Proceeding with the \textbf{Spatial Uniformity Assumption} established in Section \ref{sec:quant_comp}, we focus our analysis on the high-resolution regime. This regime is directly aligned with the requirements of scientific simulations, high-accuracy numerical modeling, and high-fidelity reconstruction, where quantization errors are expected to be much smaller than the intrinsic spatial variations of the data, making the local point distribution within each voxel well approximated as uniform. We empirically validate this assumption by our experiments shown in Figure~\ref{fig:error-distributions}, where the actual error distributions closely match the simulations derived from this uniform model. 
Then we prove that the search for an optimal quantization voxel under the MSE distortion metric is fundamentally restricted to the set of convex polyhedrons.

\begin{boxedlemma}[lemma:convexity]
Let a quantizer achieve the global minimum MSE. Then each of its quantization voxels is a convex polyhedron.
\end{boxedlemma}

\noindent\textbf{Proof.}
Let a quantizer $Q$ be defined by a set of centers $\{\mathbf{c}_i\}$ and a corresponding partition of space into voxels $\{\lambda_i\}$. The total MSE is $E(Q) = \sum_i \int_{\lambda_i} \|\mathbf{x} - \mathbf{c}_i\|^2 d\mathbf{x}$. For $E(Q)$ to be minimal for a given set of centers, the partition $\{\mathcal{R}_i\}$ must follow the nearest-neighbor rule, which means the closest center of a point $p_i$ is the center of the region $R$ that $P_i$ belongs to. We prove this by contradiction.

Let us assume that an optimal partition $\{\mathcal{R}_i^*\}$ for a given set of centers $\{\mathbf{c}_i^*\}$ is not a nearest-neighbor partition. Formally, this corresponds to:
\begin{align*}
     & \exists \mathbf{p} \in \mathcal{R}_i^* \text{ and } j \neq i \text{ such that } \|\mathbf{p} - \mathbf{c}_j^*\|^2 < \|\mathbf{p} - \mathbf{c}_i^*\|^2. \\
    \Rightarrow & \text{Construct } \{\mathcal{R}'_k\} \text{ by reassigning } \mathbf{p} \text{ from } \mathcal{R}_i^* \text{ to } \mathcal{R}_j^*. \\
    \Rightarrow & \text{The change in total MSE is }  \Delta E = \|\mathbf{p} - \mathbf{c}_j^*\|^2 - \|\mathbf{p} - \mathbf{c}_i^*\|^2 < 0.
\end{align*}
The existence of a partition with a lower total MSE contradicts the assumption that $\{\mathcal{R}_i^*\}$ is optimal. Thus, any MSE-optimal partition must be a nearest-neighbor partition.

The regions included by this nearest-neighbor rule can now be analyzed. For any two distinct centers, $\mathbf{c}_i^*$ and $\mathbf{c}_j^*$, the boundary of points equidistant from both is a hyperplane. The set of all points closer to $\mathbf{c}_i^*$ 
is formally defined by the inequality $\|\mathbf{p} - \mathbf{c}_i^*\| \le \|\mathbf{p} - \mathbf{c}_j^*\|$, which describes a closed half-space. Since a half-space is a convex set, the quantization voxel $\lambda_i^*$ can be defined as the set of all points closer to $\mathbf{c}_i^*$ than to any other center, which is therefore the intersection of multiple such half-spaces (one for each $j \neq i$).  Consequently, each voxel $\lambda_i^*$ is necessarily a convex polyhedron as the intersection of any number of convex sets is a convex set.
\hfill $\Box$

Lemma~\ref{lemma:convexity} states that the optimal quantization voxel must be a convex polyhedron. Whose representative point that minimizes the internal MSE is its centroid~\cite{Gersho1979}. 
We then formally define a criterion for comparing the efficiency of different voxel geometries.

\begin{boxedlemma}[lemma:criterion]
Consider two lattice quantizers, $Q_A$ and $Q_B$, generated by space-filling convex polyhedrons~\cite{voronoi} $\mathcal{P}_A$ and $\mathcal{P}_B$ respectively. At a constant rate (i.e., when their voxel volumes are equal), $Q_A$ achieves a lower MSE than $Q_B$ if and only if the dimensionless second moment of its voxel is lower, i.e., $G(\mathcal{P}_A) < G(\mathcal{P}_B)$.
\end{boxedlemma}

\begin{table*}[] 
\centering
\footnotesize
\begin{threeparttable}
\caption{Comparison of dimensionless second moment $G(\mathcal{P})$ for various three-dimensional polyhedra $\mathcal{P}$}
\renewcommand{\arraystretch}{1.2}
\label{tab:second-moment} 
\begin{tabular}{l | c | c | c | c | c | c | c | c} 
\hline
$P$ & Tetrahedron & \textcolor{red}{Cube$^{*}$} & Hexagonal prism$^{*}$ & Rhombic dodecahedron$^{*}$ & \textcolor{blue}{Truncated octahedron$^{*}$} & Dodecahedron & Icosahedron & Sphere \\
\hline
$G(P)$ & 0.1040042 \dots & 0.0833333 \dots & 0.0812227 \dots & 0.0787451 \dots & 0.0785433 \dots & 0.0781285 \dots & 0.0778185 \dots & 0.0769670 \dots \\
\hline
\end{tabular}
\begin{tablenotes}
\scriptsize
\item $^{*}$ A space-filling polyhedron.
\end{tablenotes}
\end{threeparttable}
\end{table*}

\noindent\textbf{Proof.}
Let $MSE_A$ and $MSE_B$ denote the MSE of the quantizers $Q_A$ and $Q_B$, respectively. The overall MSE is proportional to the unnormalized MSE within a single quantization voxel, $\mathcal{E}(\lambda)$:
$$
E(\lambda) = \frac{1}{V(\lambda)} \int_{\lambda} ||\mathbf{x} - \hat{\mathbf{x}}||^2 \,d\mathbf{x}
$$
Suppose the shape of voxel $\lambda$ is a polyhedron $\mathcal{P_\lambda}$, The dimensionless second moment, $G(\mathcal{P})$, is a scale-invariant measure of the voxel's shape efficiency, defined as:
$$
G(\mathcal{P}_\lambda) = \frac{\int_{\lambda} ||\mathbf{x} - \hat{\mathbf{x}}||^2 \,d\mathbf{x}}{[V(\lambda)]^{\frac{n+2}{n}}}
$$
By substituting the definition of $E(\lambda)$ into the formula for $G(\mathcal{P}_\lambda)$, we can express the voxel's MSE as a direct function of its shape  and volume efficiency:
$
E(\lambda) = G(\mathcal{P}_\lambda) \cdot [V(\mathcal{P}_\lambda)]^{2/n}
$.

Lemma~\ref{lemma:criterion} states that both quantizers operate at a constant rate, which implies their voxel volumes are equal. Let $V = V(\lambda_A) = V(\lambda_B)$. We can now establish a chain of logical equivalences:
\begin{align*}
    & \text{$Q_A$ achieves a lower MSE than $Q_B$} \\
    \Leftrightarrow & \mathrm{MSE}_A < \mathrm{MSE}_B \\
    \Leftrightarrow & E(\lambda_A) < E(\lambda_B) && \text{(Overall MSE = voxel MSE)} \\
    \Leftrightarrow & G(\mathcal{P}_{\lambda_A}) [V(\lambda_A)]^{2/n} \\
    & \qquad < G(\mathcal{P}_{\lambda_B}) [V(\lambda_B)]^{2/n} && \text{(By MSE relation)} \\
    \Leftrightarrow & G(\mathcal{P}_{\lambda_A}) V^{2/n} < G(\mathcal{P}_{\lambda_B}) V^{2/n} && \text{(Premise: $V(\lambda_A)\!=\!V(\lambda_B)$)} \\
    \Leftrightarrow & G(\mathcal{P}_{\lambda_A}) < G(\mathcal{P}_{\lambda_B}) && \text{(const. $V^{2/n} > 0$)}
\end{align*}
This directly shows that a lower MSE is equivalent to a lower dimensionless second moment, given equal voxel volumes.
\hfill $\Box$

Lemma~\ref{lemma:criterion} provides a definitive metric, $G(P)$, for evaluating the quantization efficiency of any given convex polyhedron. We now address whether an optimal tiling should be composed of a single uniform shape or a composite of multiple different shapes.

\begin{boxedlemma}[lemma:uniformity]
 To minimize L2-based MSE, a uniform tiling composed of a single optimal polyhedron is strictly superior to any composite tiling of varying shapes, orientations, or volumes.
\end{boxedlemma}

\noindent\textbf{Proof.}
We compare two scenarios for tiling a given space. First, consider a composite tiling of $N$ voxels with potentially varying shapes $\{S_i\}$ and volumes $\{V_i\}$. Its total squared error is $\mathcal{E}_{\text{comp}} = \sum_{i=1}^{N} G({S_i}) V_i^{5/3}$. Second, consider a uniform tiling of $N$ voxels, where each voxel has the optimal shape (i.e., with shape constant $G_{\text{opt}}$) and the same average volume $V_{\text{avg}} = (\sum V_i)/N$. Its total error is $E_{\text{uni}} = N \cdot G_{\text{opt}} \cdot (v_{\text{avg}})^{5/3}$.

We prove $E_{\text{comp}} \ge E_{\text{uni}}$ in two steps. 
By definition, the shape constant of any voxel, $G({S_i})$, cannot be better than that of the optimal shape, $G_{\text{opt}}$, thus $G({S_i}) \ge G_{\text{opt}}$. This yields the first inequality:
\[
E_{\text{comp}} = \sum_{i=1}^{N} G({S_i}) V_i^{5/3} \ge \sum_{i=1}^{N} G_{\text{opt}} V_i^{5/3} = G_{\text{opt}} \sum_{i=1}^{N} V_i^{5/3}
\]

Next, we apply \textbf{Jensen's inequality} to the convex function $f(V) = V^{5/3}$, which states that the average of the function's values is greater than or equal to the function of the average value. This gives the second inequality:
\begin{align*}
 & \frac{1}{N}\sum_{i=1}^{N} V_i^{5/3} \ge \left(\frac{\sum_{i=1}^{N} V_i}{N}\right)^{5/3} = (V_{\text{avg}})^{5/3} \Rightarrow\\
 & \sum_{i=1}^{N} V_i^{5/3} \ge N \cdot (V_{\text{avg}})^{5/3} 
\Rightarrow  \sum_{i=1}^{N} G_{\text{opt}} V_i^{5/3} \ge N \cdot G_{\text{opt}} \cdot (V_{\text{avg}})^{5/3}
\end{align*}

Combining these steps, we find that $\mathcal{E}_{\text{comp}} \ge \mathcal{E}_{\text{uni}}$. This inequality is strict in any practical composite tiling, as it will either contain suboptimal shapes (where $G({S_i}) > G_{\text{opt}}$), including rotated versions of an optimal shape, or non-uniform volumes ($V_i \neq V_{\text{avg}}$). A uniform tiling with the single optimal shape is therefore provably the most efficient solution.
\hfill $\Box$

\begin{table}[]
\centering
\footnotesize
\begin{threeparttable}
\renewcommand{\arraystretch}{1.2}
\caption{Comprehensive comparison of quantization efficiency for various space-filling polyhedra under different metrics, using the Cube as a baseline. \textbf{Bold} indicates the best choices.}
\label{tab:comprehensive-comparison}
\begin{tabular}{l | c | c c | c}
\hline
\hline
& \textbf{Intrinsic} & \multicolumn{2}{c|}{\textbf{Constant}} & \textbf{Constant} \\
& \textbf{Efficiency} & \multicolumn{2}{c|}{\textbf{MSE/PSNR}} & \textbf{Hausdorff} \\
\textbf{Polyhedron ($\mathcal{P}$)} & $G(\mathcal{P})$\tnote{a} & Volume $\Uparrow$ \tnote{b} & Voxel $\Downarrow$ \tnote{c} & Efficiency $\Uparrow$ \tnote{d} \\
\hline
\textcolor{red}{Cube} (Baseline) & 0.08333 & 1.0x & 0.0\% & 1.0x \\
Hexagonal prism & 0.08122 & 1.039x & 3.7\% & 1.54x\tnote{e} \\
\textcolor{blue}{Truncated octahedron} & \textbf{0.07854} & \textbf{1.092x} & \textbf{8.4\%} & 1.77x \\
Rhombic dodecahedron & 0.07875 & 1.088x & 8.1\% & \textbf{1.88x} \\
\hline
\hline
\end{tabular}
\begin{tablenotes}
\scriptsize
\item[a] \textbf{Dimensionless Second Moment ($G(\mathcal{P})$)}: The intrinsic measure of MSE efficiency. Lower is better.
\item[b] \textbf{Volume Gain}: Relative voxel volume for the same MSE. Higher is better.
\item[c] \textbf{Voxel Reduction}: Percentage reduction in total voxels for the same total MSE. Higher is better.
\item[d] \textbf{Hausdorff Efficiency Gain}: A proxy for performance under a Hausdorff distance constraint, based on the voxel's covering efficiency. Higher is better.
\item[e] \textbf{Note on Hexagonal Prism}: Unlike the other polyhedra, the geometry of a hexagonal prism depends on its height-to-width ratio. The value shown corresponds to the theoretical maximum efficiency,  achieved at an optimal ratio of $h/a = \sqrt{2}$.
\end{tablenotes}
\end{threeparttable}
\end{table}

Lemma~\ref{lemma:convexity}--\ref{lemma:uniformity} provides the formal underpins for narrowing the search for an MSE-optimal quantizer to a single, uniformly-tiled, space-filling convex polyhedron that minimizes the dimensionless second moment, $G(P)$. The solution to this problem is found by consulting the established values of $G(P)$ for various polyhedra in the classical 3D space-filling theory~\cite{voronoi}.

Table~\ref{tab:second-moment} shows that the Truncated Octahedron (TO), the Voronoi voxel of the Body-Centered Cubic (BCC) lattice, has the lowest known normalized second moment among 3D space-filling candidates, with $G(P)\approx0.07854$. As quantified in Table~\ref{tab:comprehensive-comparison}, replacing the cubic lattice, for which $G(P)\approx0.08333$, with the BCC lattice achieves the same MSE using \textbf{8.4\% fewer voxels}. This reduction improves rate-distortion performance because fewer voxel indices must be encoded at the same reconstruction quality. We therefore use the BCC lattice as the basis of our quantization strategy.

This geometric advantage is useful only if quantization remains computationally efficient. Although the truncated octahedron is more complex than a cube, the following section presents a fast algebraic BCC quantization algorithm with complexity comparable to standard cubic-lattice quantization.

\subsection{Efficient TO Quantization}

\begin{figure}[]
    \centering
    \includegraphics[width=\linewidth]{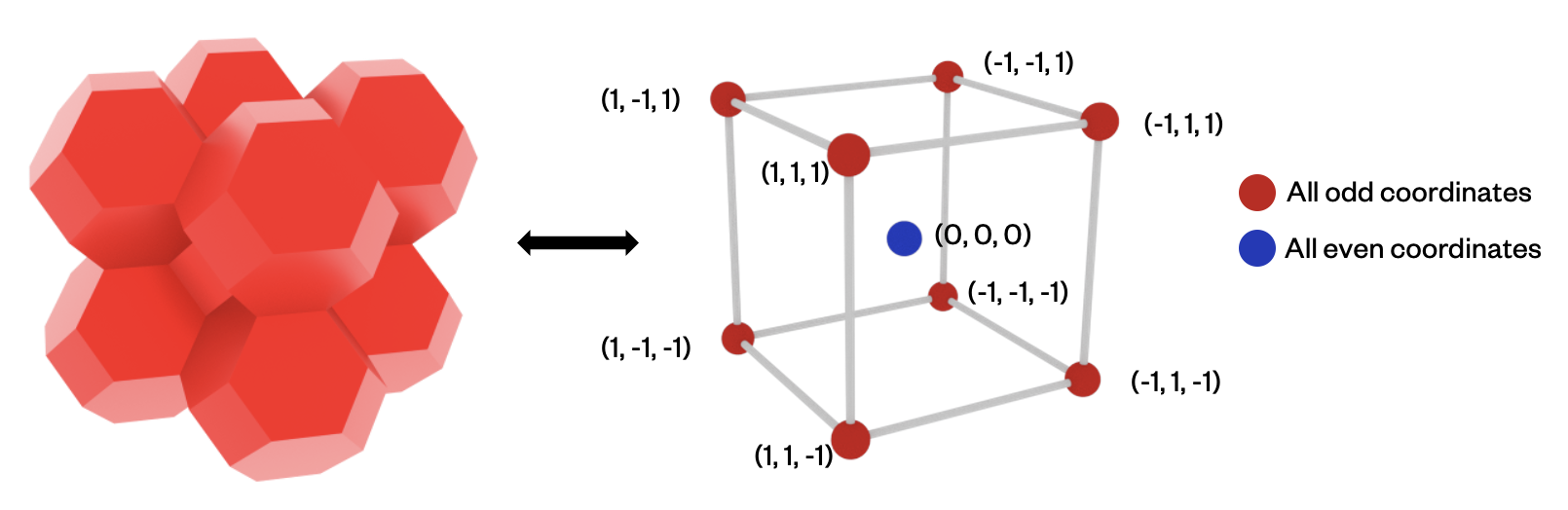}
    \caption{
        The centers of truncated octahedra form a BCC layout.
        Each center point is constructed using either all even or all odd coordinates
        in the normalized BCC coordinate system.
    }
    \label{fig:bcc-layout}
\end{figure}

Cube quantization is intuitive with low computational complexity,
as mapping a point to a cubic lattice only requires a component-wise floor or round operation. 
At first glance, adopting a more geometrically complex quantizer like the Truncated Octahedron seems to compromise this efficiency. 
However, this is not the case. The quantization can be simplified by focusing on the spatial arrangement of the voxel centers, which form a BCC lattice. For a BCC lattice aligned with an integer grid, the coordinates $(x, y, z)$ of any lattice point satisfy a simple parity property: \textbf{the components are either all even or all odd}, as shown in Figure~\ref{fig:bcc-layout}. This property allows for a direct, analytical mapping from a continuous coordinate to a discrete lattice index.

Our BCC quantization implementation has constant-time computation comparable to cube quantization. 
Given a point $\mathbf{p}$ and L2 error bound $\epsilon$, it first normalizes the point by the BCC lattice scale $2\epsilon/\sqrt{5}$. 
It then locates the unit cube containing the normalized point and identifies the two BCC lattice candidates in that cube: the all-even and all-odd integer vertices. The quantized coordinate is obtained by selecting the candidate with the smaller Euclidean distance to the normalized point. \rTwo{For example, when $\lfloor \mathbf{p}'\rfloor=(0,1,2)$, the local cube contains two valid BCC lattice candidates: the all-even point $\mathbf{q}_E=(0,2,2)$ and the all-odd point $\mathbf{q}_O=(1,1,3)$. This procedure requires a few fixed operations, yielding $\Theta(1)$ complexity while exploiting the periodic structure of the BCC lattice.}

\subsection{Encoding}
\label{sec:encoding}

After quantization, the continuous point cloud is transformed into a set of discrete integer coordinates. The remaining task is to losslessly encode these coordinates by exploiting two types of redundancy: inter-axis correlation among $(x,y,z)$ coordinates and spatial locality among nearby occupied cells.

\textbf{Joint Coordinate Encoding}. A straightforward strategy is to encode the $x$, $y$, and $z$ coordinates as three independent streams. However, for geometric data, these coordinates are inherently correlated. From an information-theoretic perspective, joint encoding is more efficient because the joint entropy $H(X,Y,Z)$ is no larger than the sum of marginal entropies $H(X)+H(Y)+H(Z)$~\cite{Cover2006}. Therefore, instead of compressing each axis independently, XnYZip first maps each 3D voxel coordinate to a single 1D index and then encodes the resulting index sequence.

\textbf{Serialization via Space-Filling Curves}. The set of quantized voxel coordinates is unordered. To exploit spatial locality, we serialize the 3D coordinates into a 1D sequence using a space-filling curve (SFC). A simple row-major order is easy to compute, but it creates large artificial jumps at row and plane boundaries, as shown in Figure~\ref{fig:sfc}. Block-wise row-major indexing, as used in compressors such as LCP, reduces the dynamic range by converting global coordinates into local coordinates, but it does not fully remove the long-tail behavior caused by row and plane transitions.

To reduce these jumps, we use locality-preserving SFCs for the 3D-to-1D projection. SFC ordering keeps nearby 3D cells close in the serialized sequence, producing smaller consecutive differences and a more compressible stream. In the evaluation, we denote the Z-order and Hilbert variants of our framework as \texttt{XnYZip-z} and \texttt{XnYZip-h}, respectively.

\textbf{Delta Encoding}. After SFC serialization, we obtain an ordered integer sequence $S$. The output of quantization is a \textbf{sparse set of discrete integer coordinates}, rather than a dense scalar field where every grid location has a value. Therefore, interpolation- or stencil-based predictors, such as Lorenzo-style predictors, are not suitable because they assume regular neighboring samples exist.

Instead, the most reliable local correlation comes from the 1D adjacency induced by SFC ordering. We therefore use a first-order predictor where the prediction for $S_i$ is its immediate predecessor $S_{i-1}$. The residual is encoded by delta encoding:
\rOne{Delta encoding is widely used in prediction-based scientific data compressors~\cite{sz3,LCP}. We adopt it here because it relies only on the local adjacency induced by SFC ordering and does not assume an underlying grid structure or functional smoothness required by interpolation- or stencil-based predictors.}

\textbf{Optional Optimization: Run-Length Encoding}. The output of quantization is an unordered multiset of $N$ integer coordinates. In dense datasets, multiple points may be quantized to the same coordinate, creating many duplicates. To exploit this redundancy, XnYZip optionally applies \textbf{run-length encoding (RLE)} immediately after quantization. RLE transforms the multiset of $N$ points into unique coordinate-count pairs, allowing the later stages to operate on a smaller set of occupied voxel locations while encoding the count stream separately.

\textbf{Entropy and Dictionary Coding}. After RLE, SFC serialization, and delta encoding, the remaining residual stream mainly contains statistical redundancy. We use \textbf{Huffman coding} to assign shorter codes to more frequent residual symbols. Finally, the resulting byte stream is passed to \textbf{Zstd} to capture remaining byte-level patterns and improve the overall compression ratio.

\rTwo{Worked Example: Encoding a 6-Cell Block.}
\rTwo{To make the bit budget concrete, consider a 13-point local block whose BCC quantization stage emits 6 unique cells with the integer coordinates and per-cell counts below. Each cell coordinate is either all-even or all-odd, as required by the BCC lattice.}

\begin{center}\footnotesize
\begin{tabular}{l|ccc}
\toprule
cell $(x,y,z)$ & count & Z-order index & delta \\
\midrule
$(0,0,0)$ & 3 & 0  & 0 \\
$(1,1,1)$ & 4 & 7  & 7 \\
$(3,3,1)$ & 1 & 15 & 8 \\
$(0,0,2)$ & 1 & 32 & 17 \\
$(1,1,3)$ & 2 & 39 & 7 \\
$(2,2,0)$ & 2 & 48 & 9 \\
\bottomrule
\end{tabular}
\end{center}

\rTwo{The RLE stage decomposes the 13-point multiset into the cell-index column and the count column above; the two streams are encoded independently from this point on. Sorting cells by their Z-order index produces a monotone sequence whose first-order differences (delta column) are bounded by the local cell spacing rather than by the bbox diagonal.}

\rTwo{On the delta stream $[0,7,8,17,7,9]$ the only repeated symbol is $7$ (frequency 2); a Huffman code that assigns $7{\to}0$, $0{\to}100$, $8{\to}101$, $17{\to}110$, $9{\to}111$ encodes the six symbols in $3{+}1{+}3{+}3{+}1{+}3=14$ bits. On the count stream $[3,4,1,1,2,2]$, with frequencies $\{1{:}2$, $2{:}2$, $3{:}1$, $4{:}1\}$, a Huffman code $2{\to}0,\ 1{\to}10,\ 3{\to}110,\ 4{\to}111$ uses $12$ bits. The whole block is therefore encoded in $14+12=26$ bits of payload, compared with $54$ bits for the raw cell-and-count list ($6\times 6$-bit Morton codes plus $6\times 3$-bit counts) and $\approx 78$ bits if the 13 original points were written as integer triples. The downstream Zstd stage further compresses the byte-level structure of the Huffman output.}

\subsection{Distributed Compression}
\label{sec:distributed}

\rOne{The single-node pipeline in Section~\ref{sec:encoding} assumes that the input point cloud can be processed within the memory of one machine. This assumption is often violated by production scientific workloads. For example, a single HACC particle file contains on the order of $10^{9}$ points (about $12.9$~GB), and a full simulation snapshot is partitioned into hundreds of such files. To support such workloads, we extend XnYZip to an MPI-based map-shuffle-reduce pipeline over the quantized cell table, as shown in Algorithm~\ref{alg:mapreduce}.}

\rOne{Each rank reads a disjoint slice of the input, computes the global bounding box using one \texttt{MPI\_Allreduce}, and quantizes its local points into a cell-to-count hash map (Map). The ranks then route cell-count entries according to their corresponding SFC key ranges using \texttt{MPI\_Alltoallv} (Shuffle). After the shuffle, each rank owns a disjoint contiguous SFC range, aggregates duplicate cells received from different ranks (Reduce), encodes the owned cells using Section~\ref{sec:encoding}, and writes a self-contained part file. This representation preserves the same global cell-count table as the monolithic compressor: each occupied cell appears in exactly one rank's output. The only extra storage comes from per-rank headers and part boundaries. Moreover, the shuffle operates on unique cell-count pairs rather than raw points, so dense point clouds often incur much lower communication volume than their input size.}

\begin{algorithm}[]
\footnotesize
\caption{Distributed Compression via MPI Map-Shuffle-Reduce}
\label{alg:mapreduce}
\SetAlgoLined
\DontPrintSemicolon
\SetKwInOut{KwIn}{Input}
\SetKwInOut{KwOut}{Output}

\KwIn{Point cloud file $F$, error bound $\epsilon$, rank $r$ out of $R$ ranks}
\KwOut{Compressed file.}

$slice_r \leftarrow$ byte range of $F$ assigned to rank $r$;
$(\min, \max) \leftarrow$ \texttt{MPI\_Allreduce}(per-axis extrema of $slice_r$);
$local\_table \leftarrow$ empty hash map;

\ForEach{point $\mathbf{p} \in slice_r$}{
$cell \leftarrow$ BCCQuantize$(\mathbf{p}, \epsilon)$;
$local\_table[cell] \mathrel{+}= 1$;
}

\ForEach{$(cell, count) \in local\_table$}{
$key \leftarrow \textsc{SFC}(cell)$;
append $(cell, count)$ to $send_buf[\textsc{owner}(key)]$;
}

$recv_buf \leftarrow$ \texttt{MPI\_Alltoallv}$(send_buf)$;
$owned \leftarrow$ aggregate counts in $recv_buf$ by $cell$;
$bytes \leftarrow$ Encoding pipeline of S~\ref{sec:encoding} applied to $owned$;
\end{algorithm}

\rOne{A fixed-bit-width Morton partition can assign SFC ranges to ranks with low overhead, but it may suffer from load imbalance when the bounding box is highly skewed. For example, on the HACC volume of $32 \times 64 \times 256$, the longest axis dominates the leading Morton bits and can concentrate many cells onto a few ranks. We therefore use sample-sort splitters: each rank samples local cells, rank 0 selects approximate $(i/R)$-th quantile splitters, and the splitters are broadcast to all ranks. The encoding pipeline is unchanged, but the SFC ranges are chosen to balance the number of owned cells across ranks.}

\rOne{This design follows the parallelization profile of the single-node pipeline. TO quantization and local hash-table construction are embarrassingly parallel, and each rank performs local sorting and encoding independently after the shuffle. Since point clouds are unordered in our target workloads, the original input-file order is not preserved across cells; order-sensitive workloads would require an additional global-index stream during decompression.}

\rOne{On a 3GB HACC subset compressed with 4 ranks at $eb=10^{-3}$, the gather-and-expand decompressor recovers every point within the prescribed bound.  On a full $12.9$~GB HACC file with $1.07 \times 10^{9}$ points, it sustains above $80\%$ iso efficiency from 1 to 16 nodes for compute and encoding phases, while the shuffle accounts for less than $4\%$ of the total wall-clock time. The distributed implementation retains $99.7\%$ of the monolithic compression ratio at 4 ranks and $99.5\%$ at 16 ranks on both subset and fullset of HACC.}

\section{Evaluation}
\label{sec:eval}

We evaluate XnYZip on 8 point cloud datasets against \rOne{7} state-of-the-art baselines. The experiments cover four aspects: theory-practice consistency, end-to-end performance,
throughput, and case studies. The datasets include scientific and non-scientific domains, with both dense simulation data and conventional point clouds.

\subsection{Experiment Setup}
\noindent\textbf{Datasets}. We evaluate our framework on eight datasets from diverse scientific and industrial domains, listed in Table~\ref{tab:datasets}. Our evaluation focuses on the 3D spatial coordinates $(x,y,z)$ for all datasets. For the original terabyte-scale datasets (EXAALT, USGS, HACC, WarpX) and other large-scale data, we extract representative subsets for evaluation. Smaller datasets, such as Stanford Dragon and GroEL-GroES (1AON), are compressed in their entirety. For datasets with a temporal dimension, such as YIIP and WarpX, we select one representative frame because our work focuses on compressing individual point-cloud snapshots.

\noindent\textbf{Hardware \& Software}. All experiments are conducted on Bebop Cluster of ANL. Each compute node has dual-socket Intel Xeon E5-2695 v4 processors with 36 physical cores in total and 128 GB DDR4 memory. Our code is compiled using GCC 13.2.0 with the \texttt{-O3} optimization flag.

\begin{table}[]
\centering
\scriptsize
\setlength{\tabcolsep}{3pt}
\caption{Benchmark datasets used in the evaluation. \emph{Size} denotes the size of the upstream corpus from which each dataset is drawn; \protect\rOne{\emph{Tested Size} and \emph{Points} denote the data actually compressed in our experiments, which for terabyte-scale corpora corresponds to representative subsets of the full archive.}}
\label{tab:datasets}
\begin{tabular}{l l r r r r}
\hline
\textbf{Dataset} & \textbf{Domain} & \textbf{Size} & \rOne{\textbf{Tested Size}} & \rOne{\textbf{Points}} & \rOne{\textbf{$\rho^{\ddagger}$}} \\
\hline
YIIP~\cite{YIIP}                       & MD$^{\ast}$      & 4 GB      & \rOne{1.3 MB}              & \rOne{$1.12{\times}10^{5}$} & \rOne{1.00} \\
Stanford Dragon~\cite{stanford3dscan}  & CG$^{\ast}$      & 43 MB     & \rOne{5.0 MB}              & \rOne{$4.38{\times}10^{5}$} & \rOne{1.33} \\
GroEL-GroES~\cite{Xu1997}              & Biology          & 5 MB      & \rOne{0.7 MB}              & \rOne{$5.87{\times}10^{4}$} & \rOne{1.00} \\
Vesicles~\cite{Kenney2015}             & Biology          & 807 MB    & \rOne{20.0 MB}             & \rOne{$1.75{\times}10^{6}$} & \rOne{1.00} \\
USGS~\cite{usgs3dep2024}               & Geology          & $>$200 TB & \rOne{175 MB}              & \rOne{$1.53{\times}10^{7}$} & \rOne{\textbf{10.20}} \\
EXAALT~\cite{EXAALT2021}               & MD$^{\ast}$      & 4 GB      & \rOne{32.8 MB}             & \rOne{$2.87{\times}10^{6}$} & \rOne{1.00} \\
HACC~\cite{HACC}                       & Cosmology        & $>$200 TB & \rOne{12.9 GB$^{\dagger}$} & \rOne{$1.07{\times}10^{9}{}^{\dagger}$} & \rOne{\textbf{5.39}} \\
WarpX~\cite{warpx}                     & Plasma           & 8 TB      & \rOne{2.76 GB}             & \rOne{$2.41{\times}10^{8}$} & \rOne{1.01} \\
\hline
\multicolumn{6}{l}{\rOne{\scriptsize $^{\ast}$ MD for Molecular Dynamics, CG for Computer Graphics.}}\\
\multicolumn{6}{l}{\rOne{\scriptsize $^{\dagger}$ A smaller subset ($\sim$3.13\,GB) has been used for compression ratio and case study evaluation. }} \\
\multicolumn{6}{l}{\rOne{\scriptsize Larger fullset is used for distributed compression experiments. }} \\
\multicolumn{6}{l}{\rOne{\scriptsize $^{\ddagger}$ Point density $\rho = N / M_{occ}$ measured at $eb = 10^{-3}$}} \\
\hline
\end{tabular}
\end{table}

\noindent\textbf{Baselines}. We compare XnYZip with \rOne{seven} compressors covering scientific error-bounded compression, point cloud compression, and \rThree{database-side floating-point codecs}. Table~\ref{tab:baselines} summarizes their target domains and how they are used in our evaluation.

\begin{table}[]
\centering
\footnotesize
\caption{Baseline compressors used in the evaluation.}
\label{tab:baselines}
\begin{tabular}{l l l}
\hline
\textbf{Baseline} & \textbf{Type} & \textbf{Target Domain} \\
\hline
SZ3~\cite{sz3}                 & Error-bounded lossy              & Scientific arrays                \\
ZFP~\cite{zfp}                 & Error-bounded lossy              & Scientific arrays                   \\
LCP~\cite{LCP}                 & Error-bounded lossy              & Scientific particles               \\
Draco~\cite{google_draco}      & Lossy point cloud                & Graphics / media                    \\
TMC13~\cite{mpeg_pcc_tmc13}    & Lossy point cloud                & Graphics / media                       \\
\rThree{ALP~\cite{alp}}        & \rThree{Lossless float codec}    & \rThree{Numeric columns}    \\
\rThree{BUFF~\cite{buff}}      & \rThree{Bounded-precision float codec} & \rThree{Numeric columns}  \\
\hline
\end{tabular}
\end{table}

\subsection{Theory-Practice Consistency Analysis}
\label{sec:theory-practice}

\begin{figure}[]
    \centering
    \includegraphics[width=0.9\linewidth]{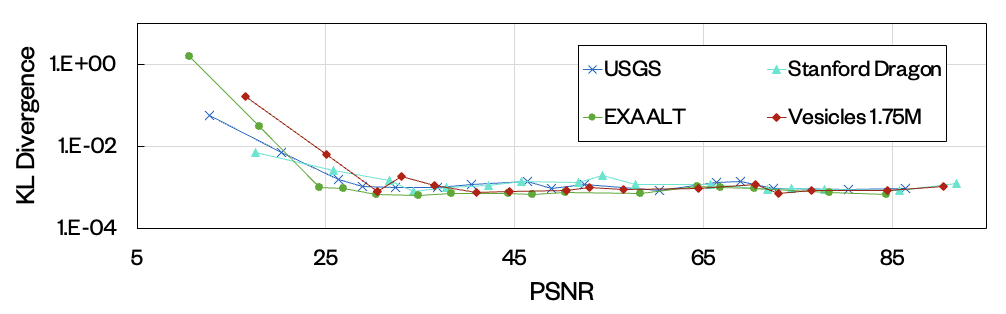}
    \caption{
    Kullback-Leibler divergence  measuring deviation between empirical error distributions and the uniform spatial error assumption across PSNR levels.
    }
    \label{fig:kl-psnr}
\end{figure}

\begin{figure}[]
    \centering
    \begin{subfigure}[]{0.45\linewidth}
        \centering
        \includegraphics[width=\linewidth]{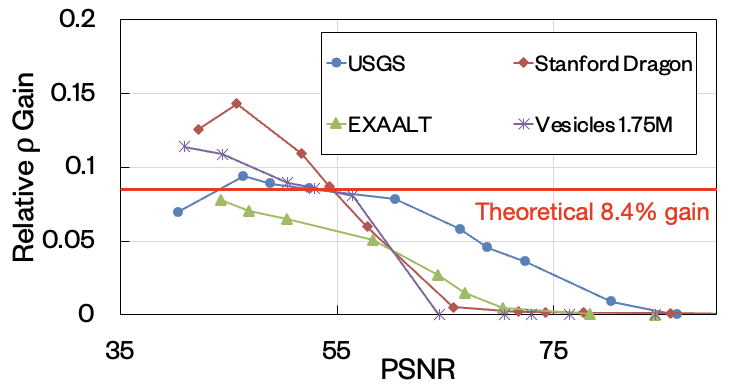}
        \caption{Relative density gain, TO over cube}
        \label{fig:rho-gain}
    \end{subfigure}
    \hfill
    \begin{subfigure}[]{0.45\linewidth}
        \centering
        \includegraphics[width=\linewidth]{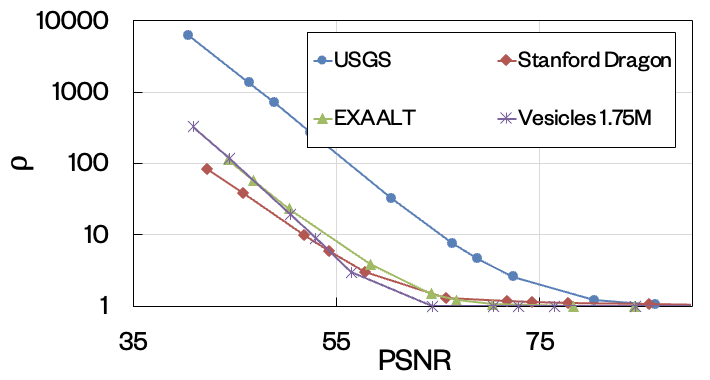}
        \caption{Density at different PSNR levels}
        \label{fig:rho-psnr}
    \end{subfigure}
    \caption{
    Geometric efficiency of Truncated Octahedron (TO) quantization across different density regimes and datasets.
    }
    \label{fig:rho-analysis}
\end{figure}

We validate three theoretical claims of Sections~\ref{sec:problemformulation}--\ref{sec:Methodology}.
(i)~The spatial-uniformity assumption underlying our quantization analysis predicts that the empirical per-voxel error distribution matches the one induced by uniformly distributed points within each voxel; Figure~\ref{fig:kl-psnr} measures the gap as a KL divergence, which converges to $\mathcal{O}(10^{-3})$ on every dataset as the bound tightens.
(ii)~The $8.4\%$ theoretical voxel-count advantage of TO over the cube in the high-density regime (Table~\ref{tab:comprehensive-comparison}) is recovered empirically in Figure~\ref{fig:rho-analysis}: a stable $\approx 8.6\%$ density reduction in the $\rho \gg 1$ regime, smoothly diminishing as $\rho \to 1$.
(iii)~The information-theoretic analysis on USGS (Table~\ref{tab:entropy-analysis}) confirms that the Hilbert curve attains the lowest post-delta residual entropy across the row-major, Z-order, and Hilbert candidates, validating our SFC family.

\begin{table}[]
\centering
\footnotesize
\caption{Information-theoretic analysis of different serialization methods on the USGS dataset at $eb=1$ (PSNR = 66\,dB). Hilbert attains the lowest post-delta residual entropy $H(R)$.}
\label{tab:entropy-analysis}
\begin{tabular}{lccc}
\toprule
\textbf{Serialization} & \multicolumn{1}{c}{\textbf{Pre-Delta}} & \multicolumn{2}{c}{\textbf{Post-Delta}} \\
\cmidrule(lr){2-2} \cmidrule(lr){3-4}
\textbf{Method} & \textbf{Cond. Entropy} & \textbf{Mutual Info} & \textbf{Entropy} \\
& $H(S_i|S_{i-1})$ & $I(R_i;R_{i-1})$ & $H(R)$ \\
\midrule
Row major order     & \textbf{0.5314} & 0.0082 & 0.8719 \\
Z-order Curve   & 0.5364 & 0.0070 & 0.8530 \\
Hilbert Curve   & 0.5324 & \textbf{0.0067} & \textbf{0.8351} \\
\bottomrule
\end{tabular}
\end{table}

\subsection{Performance Analysis}
\label{sec:perf-analysis}

Next, we evaluate \texttt{XnYZip}'s end-to-end behaviour against the baselines on the same eight datasets in different metrics.

\begin{figure}[]
    \centering
    \begin{subfigure}[]{0.4\linewidth}
        \centering
        \includegraphics[width=\linewidth]{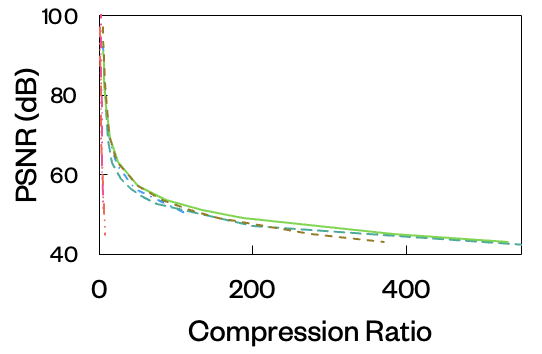}
        \caption{Stanford Dragon}
        \label{fig:rd-dragon}
    \end{subfigure}
    \hfill
    \begin{subfigure}[]{0.4\linewidth}
        \centering
        \includegraphics[width=\linewidth]{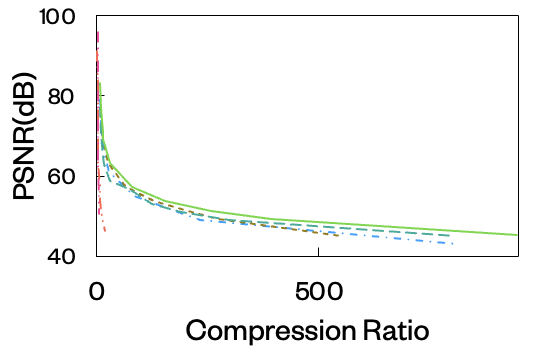}
        \caption{Vesicles}
        \label{fig:rd-vesicles}
    \end{subfigure}
    \begin{subfigure}[]{0.4\linewidth}
        \centering
        \includegraphics[width=\linewidth]{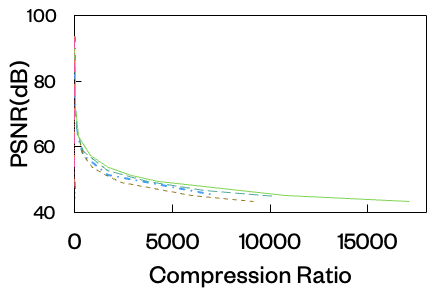}
        \caption{USGS}
        \label{fig:rd-usgs}
    \end{subfigure}
    \hfill
    \begin{subfigure}[]{0.4\linewidth}
        \centering
        \includegraphics[width=\linewidth]{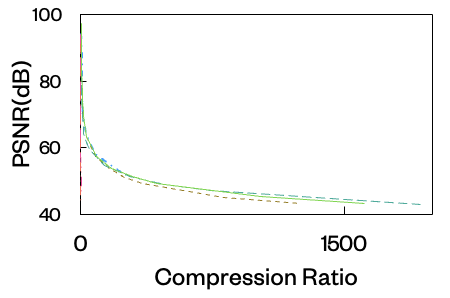}
        \caption{EXAALT}
        \label{fig:rd-exaalt}
    \end{subfigure}
    \begin{subfigure}[]{0.4\linewidth}
        \centering
        \includegraphics[width=\linewidth]{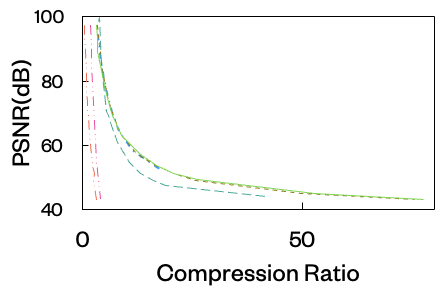}
        \caption{YIIP}
        \label{fig:rd-yiip}
    \end{subfigure}
    \hfill
    \begin{subfigure}[]{0.4\linewidth}
        \centering
        \includegraphics[width=\linewidth]{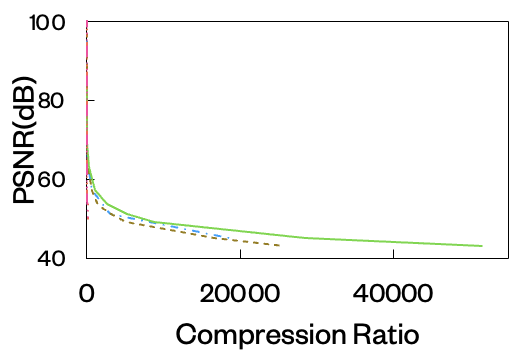}
        \caption{HACC}
        \label{fig:rd-hacc}
    \end{subfigure}
    \begin{subfigure}[]{0.4\linewidth}
        \centering
        \includegraphics[width=\linewidth]{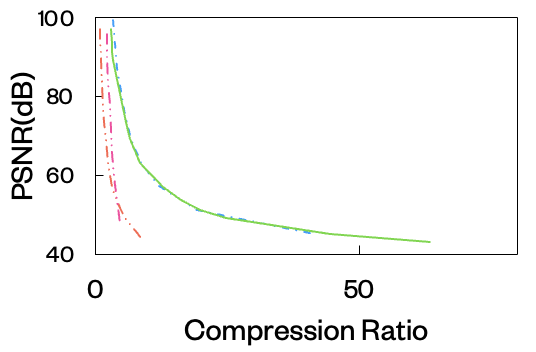}
        \caption{GroEL-GroES}
        \label{fig:rd-1aon}
    \end{subfigure}
    \hfill
    \begin{subfigure}[]{0.4\linewidth}
        \centering
        \includegraphics[width=\linewidth]{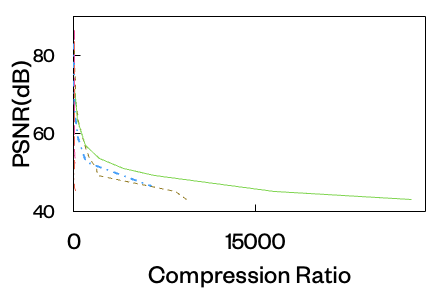}
        \caption{WarpX}
        \label{fig:rd-warpx}
    \end{subfigure}
    \begin{subfigure}[]{0.85\linewidth}
        \centering
        \includegraphics[width=\linewidth]{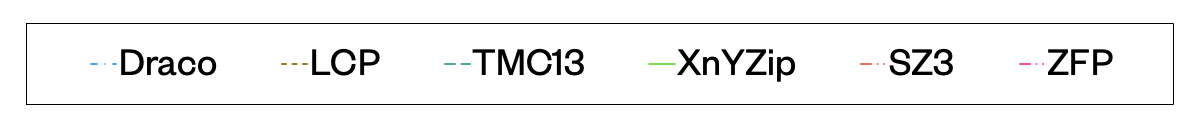}
        \label{fig:legend}
    \end{subfigure}
    \caption{
    Rate--distortion. \texttt{XnYZip} is at or above the Pareto frontier on every dataset and dominates in the low-PSNR (dense) regime.
    }
    \label{fig:rd-all}
\end{figure}

\textbf{Rate-distortion.}
Figure~\ref{fig:rd-all} shows the distortion rate per--dataset in all eight evaluation datasets. \texttt{XnYZip} lies at or above the Pareto frontier on every dataset and dominates the low-PSNR (dense) regime predicted by the theory in Section~\ref{sec:Methodology}; the advantage diminishes smoothly as occupancy thins, exactly as the $\rho$ model.

\begin{table*}[]
    \caption{Master compression-ratio table under the standard error-bounded rate--distortion convention. Each cell reports the best measured CR that satisfies the corresponding relative-L2 max-error target. A dash means the method's tightest measured configuration still exceeds the target or met bugs. The ALP row reports the lossless ALP result for reference.}
    \label{tab:appendix-master}
    \scriptsize
    \setlength{\tabcolsep}{3pt}
    \rOne{\begin{tabular}{c|cccccc||cccccc||cccccc||cccccc}
    \hline \hline
     & \multicolumn{6}{c||}{\textbf{YIIP}} & \multicolumn{6}{c||}{\textbf{Stanford Dragon}} & \multicolumn{6}{c||}{\textbf{GroEL-GroES}} & \multicolumn{6}{c}{\textbf{Vesicles}} \\ \hline
    Error & 1e-5 & 5e-5 & 1e-4 & 5e-4 & 1e-3 & 5e-3 & 1e-5 & 5e-5 & 1e-4 & 5e-4 & 1e-3 & 5e-3 & 1e-5 & 5e-5 & 1e-4 & 5e-4 & 1e-3 & 5e-3 & 1e-5 & 5e-5 & 1e-4 & 5e-4 & 1e-3 & 5e-3 \\ \hline \hline
    SZ3 & 0.62 & 0.62 & 0.78 & 1.30 & 1.60 & 2.40 & 0.97 & 1.36 & 1.62 & 2.55 & 3.26 & 6.02 & 0.68 & 1.03 & 1.21 & 1.83 & 2.28 & 4.78 & 1.09 & 1.45 & 1.68 & 2.67 & 3.56 & 10.58 \\ \hline
    ZFP & 1.54 & 1.80 & 1.91 & 2.17 & 2.33 & 2.72 & 1.79 & 2.13 & 2.27 & 2.63 & 2.86 & 3.44 & 1.69 & 2.01 & 2.14 & 2.47 & 2.68 & 3.22 & 1.77 & 2.13 & 2.28 & 2.66 & 2.90 & 3.53 \\ \hline
    LCP & \textbf{3.05} & \textbf{3.92} & \textbf{4.47} & 6.59 & 8.30 & 21.62 & \textbf{4.19} & \textbf{5.67} & 6.61 & 12.99 & 20.17 & 124.3 & \textbf{3.02} & 3.85 & 4.37 & 6.50 & 8.17 & 20.45 & \textbf{3.90} & 5.48 & 6.61 & 12.23 & 21.40 & 219.2 \\ \hline
    Draco & 2.71 & 3.64 & 4.11 & 5.52 & 6.68 & 11.47 & 3.49 & 5.18 & 6.16 & 9.82 & 13.80 & 46.74 & 3.00 & \textbf{4.17} & \textbf{4.79} & \textbf{6.84} & \textbf{8.71} & 19.21 & \textbf{3.90} & \textbf{6.13} & \textbf{7.60} & 14.63 & 28.14 & 231.6 \\ \hline
    TMC13 & – & – & – & – & 3.81 & 3.81 & – & – & – & – & – & – & – & – & 4.47 & 6.24 & 8.47 & 18.20 & 3.63 & 5.55 & 5.85 & 13.19 & \textbf{29.08} & 189.5 \\ \hline
    BUFF & 1.68 & 2.13 & 2.13 & 2.59 & 2.59 & 2.59 & 1.71 & 1.71 & 2.18 & 2.59 & 2.59 & 3.84 & 1.88 & 2.00 & 2.00 & 2.00 & 2.00 & 2.00 & 1.71 & 1.71 & 1.71 & 1.71 & 1.71 & 1.71 \\ \hline
    ALP & \multicolumn{6}{c||}{2.34 (lossless)} & \multicolumn{6}{c||}{1.32 (lossless)} & \multicolumn{6}{c||}{1.68 (lossless)} & \multicolumn{6}{c}{1.14 (lossless)} \\ \hline
    XnYZip-h & 2.81 & 3.43 & 4.33 & \textbf{6.79} & \textbf{9.00} & \textbf{26.15} & 3.29 & 5.51 & \textbf{6.69} & \textbf{14.20} & \textbf{23.76} & \textbf{187.6} & 2.74 & 3.31 & 4.20 & 6.56 & 8.55 & \textbf{24.78} & 3.55 & 5.63 & 6.85 & \textbf{15.09} & 28.73 & \textbf{391.1} \\ \hline
    XnYZip-z & 2.82 & 3.43 & 4.33 & 6.78 & 8.96 & 25.86 & 3.29 & 5.51 & \textbf{6.69} & 13.92 & 22.41 & 180.2 & 2.74 & 3.31 & 4.21 & 6.45 & 8.65 & 24.17 & 3.55 & 5.63 & 6.80 & 14.73 & 27.84 & 373.6 \\ \hline
    \hline
     & \multicolumn{6}{c||}{\textbf{USGS}} & \multicolumn{6}{c||}{\textbf{EXAALT}} & \multicolumn{6}{c||}{\textbf{HACC}} & \multicolumn{6}{c}{\textbf{WarpX}} \\ \hline
    Error & 1e-5 & 5e-5 & 1e-4 & 5e-4 & 1e-3 & 5e-3 & 1e-5 & 5e-5 & 1e-4 & 5e-4 & 1e-3 & 5e-3 & 1e-5 & 5e-5 & 1e-4 & 5e-4 & 1e-3 & 5e-3 & 1e-5 & 5e-5 & 1e-4 & 5e-4 & 1e-3 & 5e-3 \\ \hline \hline
    SZ3 & 1.25 & 1.52 & 1.78 & 2.80 & 3.75 & 7.65 & 0.75 & 0.92 & 1.02 & 1.31 & 1.49 & 2.23 & 1.46 & 2.13 & 2.66 & 5.87 & 9.81 & 84.34 & 2.52 & 3.30 & 3.87 & 5.58 & 6.89 & 14.94 \\ \hline
    ZFP & 1.84 & 2.08 & 2.22 & 2.57 & 2.79 & 3.66 & 1.63 & 1.82 & 1.93 & 2.36 & 2.54 & 3.03 & 1.97 & 2.25 & 2.42 & 2.85 & 3.12 & 4.19 & 2.09 & 2.48 & 2.63 & 2.99 & 3.21 & 3.76 \\ \hline
    LCP & 5.40 & \textbf{9.14} & 13.05 & 54.31 & 137.6 & 1685.0 & \textbf{4.06} & 5.78 & 7.06 & 15.17 & 27.53 & 265.7 & – & 11.02 & 16.10 & 62.50 & 160.8 & 3406.0 & \textbf{10.86} & \textbf{22.84} & \textbf{30.83} & \textbf{111.5} & 272.4 & 1956.9 \\ \hline
    Draco & – & 7.85 & 10.40 & 27.15 & 59.68 & 436.7 & 3.58 & 5.40 & 6.50 & 10.94 & 16.61 & 85.38 & \textbf{6.37} & \textbf{14.63} & \textbf{22.28} & 76.97 & 209.9 & 2968.2 & 4.06 & 6.35 & 7.82 & 32.95 & 88.39 & 968.37 \\ \hline
    TMC13 & – & – & – & – & – & – & – & – & – & – & – & – & – & – & – & – & – & – & – & – & – & – & – & – \\ \hline
    BUFF & 1.75 & 1.75 & 1.75 & 1.75 & 1.75 & 1.75 & 1.88 & 2.13 & 2.13 & 2.13 & 2.13 & 2.13 & 1.85 & 2.09 & 2.09 & 2.09 & 2.09 & 2.09 & 1.60 & 2.13 & 2.29 & 2.29 & 2.29 & 3.56 \\ \hline
    ALP & \multicolumn{6}{c||}{1.40 (lossless)} & \multicolumn{6}{c||}{1.18 (lossless)} & \multicolumn{6}{c||}{1.15 (lossless)} & \multicolumn{6}{c}{1.22 (lossless)} \\ \hline
    XnYZip-h & 5.44 & 6.63 & \textbf{14.55} & \textbf{69.78} & \textbf{210.6} & 4080.0 & 3.81 & \textbf{5.87} & 7.27 & \textbf{17.68} & \textbf{36.25} & \textbf{434.6} & 6.25 & 12.57 & 19.15 & \textbf{84.90} & \textbf{248.6} & 8778.6 & 5.22 & 7.63 & 10.66 & 80.70 & 327 & 6603 \\ \hline
    XnYZip-z & \textbf{5.46} & 6.65 & 14.48 & 68.38 & 207.9 & \textbf{4112.3} & 3.81 & \textbf{5.87} & \textbf{7.28} & 17.67 & 36.03 & 429.1 & 6.22 & 12.45 & 18.95 & 84.03 & 247.4 & \textbf{8894.2} & 5.70 & 11.94 & 13.26 & 102.0 & \textbf{336} & \textbf{6662} \\ \hline
    \hline
    \end{tabular}}
\end{table*}

Besides PSNR, another key distortion metric is the user-specified L2 error bound. \rOne{Table~\ref{tab:appendix-master} reports the compression ratios of eight methods on eight datasets at six relative-L2 maximum-error targets, where the L2 error is normalized by the bounding-box diagonal. Each cell gives the highest compression ratio directly measured while satisfying $|p-p'|_2/\mathrm{diag}\leq\text{target}$. Thus, entries in the same column share the same physical error bound and are directly comparable. A dash indicates that even the method's tightest measured configuration exceeds the target. BUFF is evaluated at the same L2 targets as a bounded-precision lossy baseline; because its valid bit-precision range is narrow, its ratios remain near $1\times$ and far below those of dedicated geometric codecs. Since ALP is lossless, Table~\ref{tab:appendix-master} reports one compression ratio per dataset.}

\rOne{To assess the effect of scale, we apply strided subsampling to two real-world datasets, retaining every $k$-th point for $k\in{32,16,8,4,2,1}$ to reduce density while preserving spatial extent. Compression ratio increases monotonically with scale at every error bound, showing that subsampling does not weaken our conclusion. For HACC-small, it rises from 23.0$\times$ to 93.3$\times$ at $\mathit{eb}=0.2$, while for USGS it rises from 37.4$\times$ to 382.3$\times$ at $\mathit{eb}=2$. The same trend holds at all other tested bounds, and all 72 configurations show monotone non-decreasing compression ratios as scale increases.}

\begin{figure}[]
    \centering
    \includegraphics[width=0.85\linewidth]{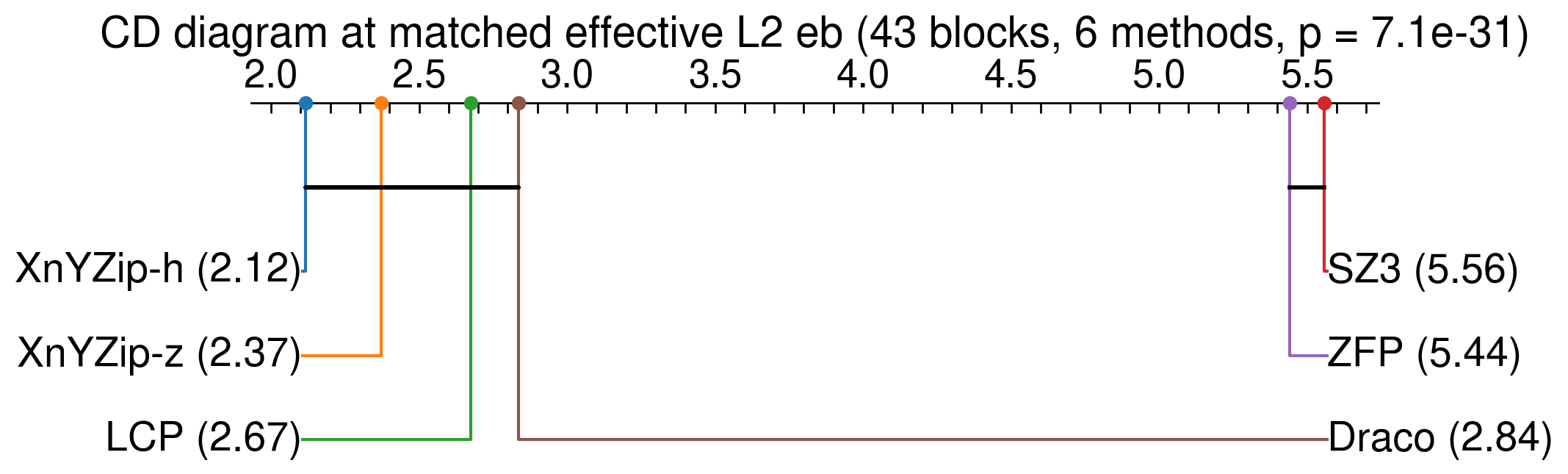}
    \caption{\rOne{Critical-difference diagram for the six lossy compressors evaluated at matched effective L2 max-error across 43 (dataset, eb) blocks spanning 8 datasets. The Friedman test yields $p = 7.1\!\times\!10^{-31}$. Methods connected by a horizontal bar are not statistically distinguishable at $\alpha = 0.05$ under the Nemenyi post-hoc test. TMC13 is excluded because its leaves too many invalid entries across the matched-error blocks; BUFF is excluded because its valid bit-precision range is too narrow to populate sufficient blocks.}}    \label{fig:cd}
\end{figure}

\rOne{\textbf{Statistical significance.}
A Friedman omnibus test followed by a Nemenyi post-hoc analysis on the seven lossy compressors at various L2 max-error (Figure~\ref{fig:cd}) places XnYZip-h/-z, Draco, and LCP in a top cluster statistically separated from ZFP and SZ3. 
}

\label{sec:cd}

\begin{figure}[]
    \centering
    \begin{subfigure}{0.3\linewidth}
        \includegraphics[width=\textwidth]{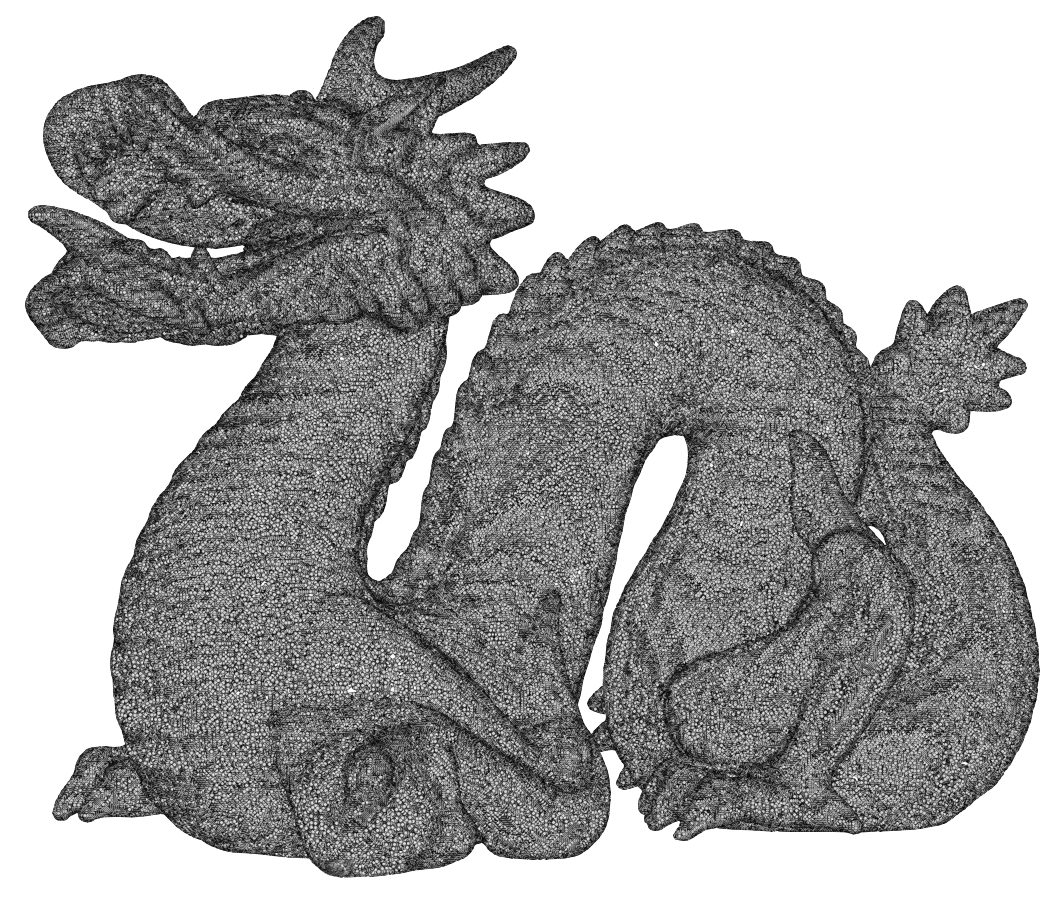}
        \subcaption{Original \\~}
        \vspace{0.9em}
        \label{fig:original-dragon}
    \end{subfigure}
    \begin{subfigure}{0.3\linewidth}
        \includegraphics[width=\textwidth]{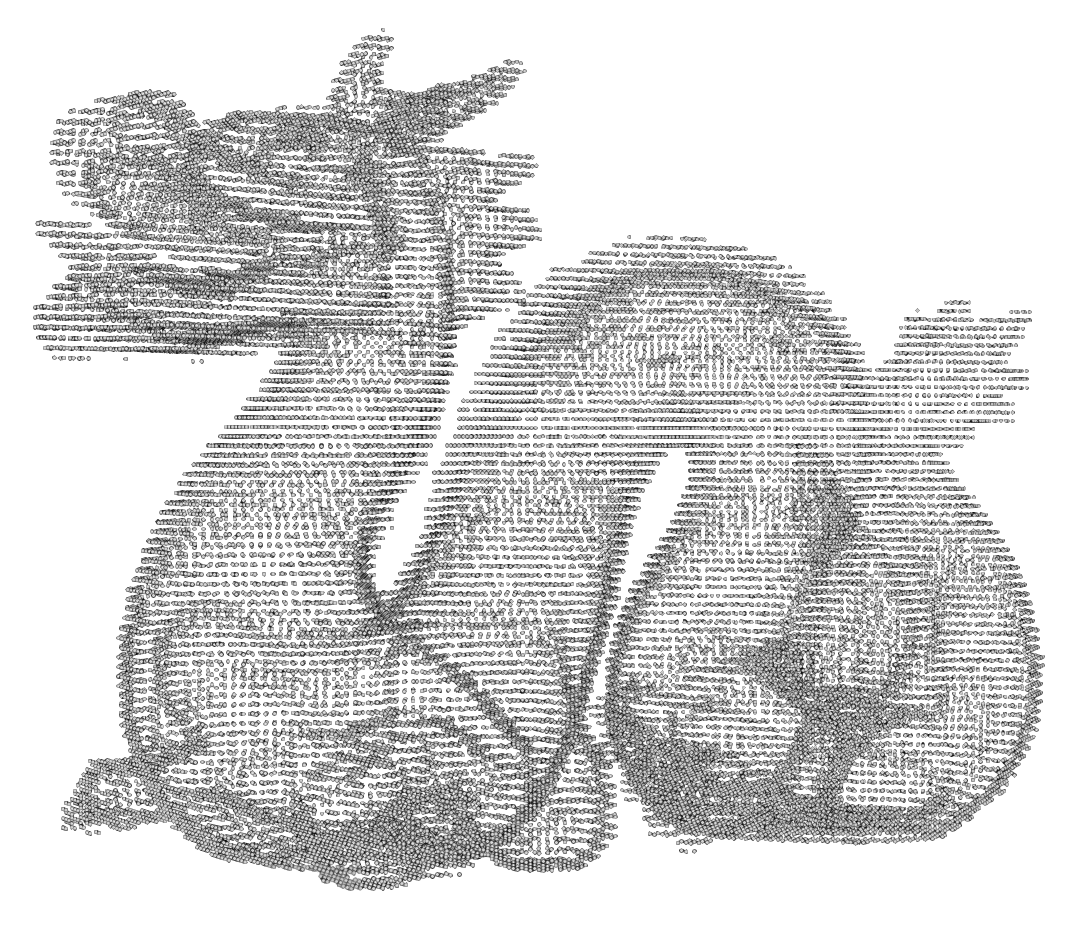}
        \subcaption{Draco \\ (CR=110.99, PSNR=50.39)}
        \label{fig:draco-dragon}
    \end{subfigure}
    \begin{subfigure}{0.3\linewidth}
        \includegraphics[width=\textwidth]{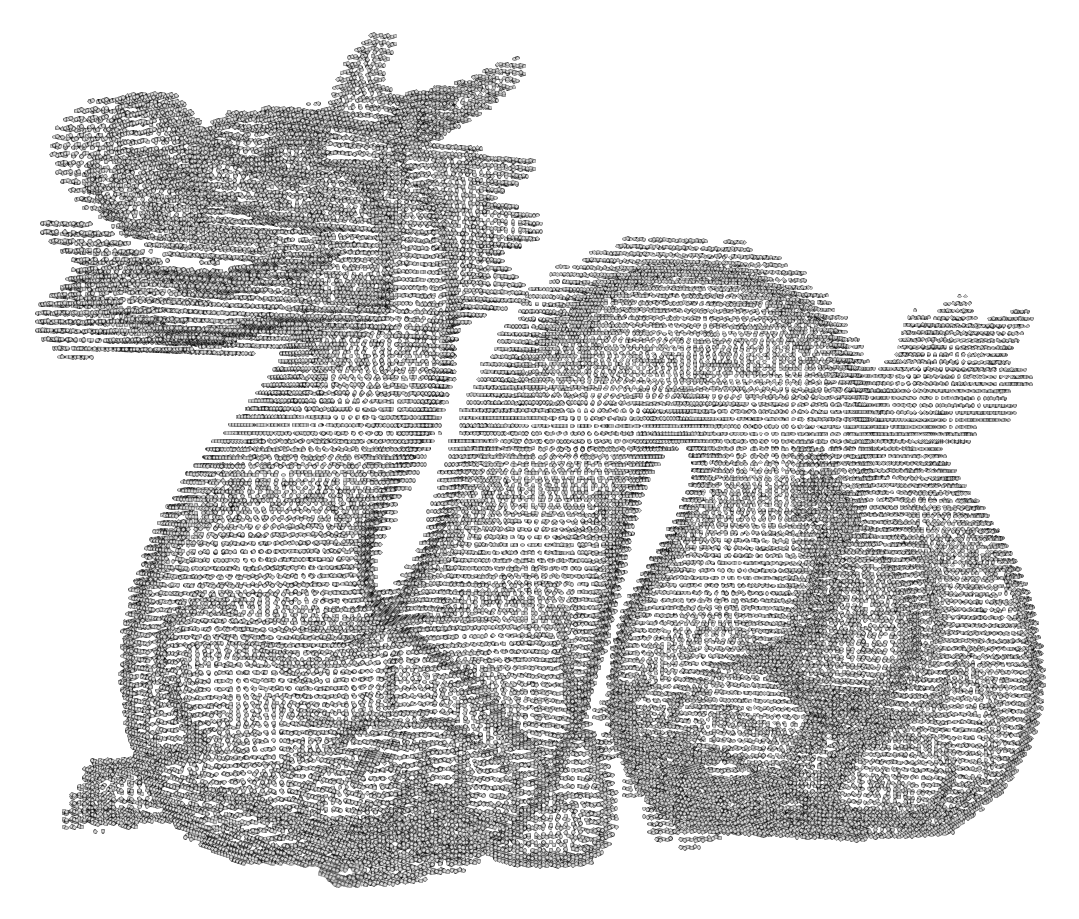}
        \subcaption{LCP \\ (CR=110.21, PSNR= 51.7)}
        \label{fig:lcp-dragon}
    \end{subfigure}
    \begin{subfigure}{0.3\linewidth}
        \includegraphics[width=\textwidth]{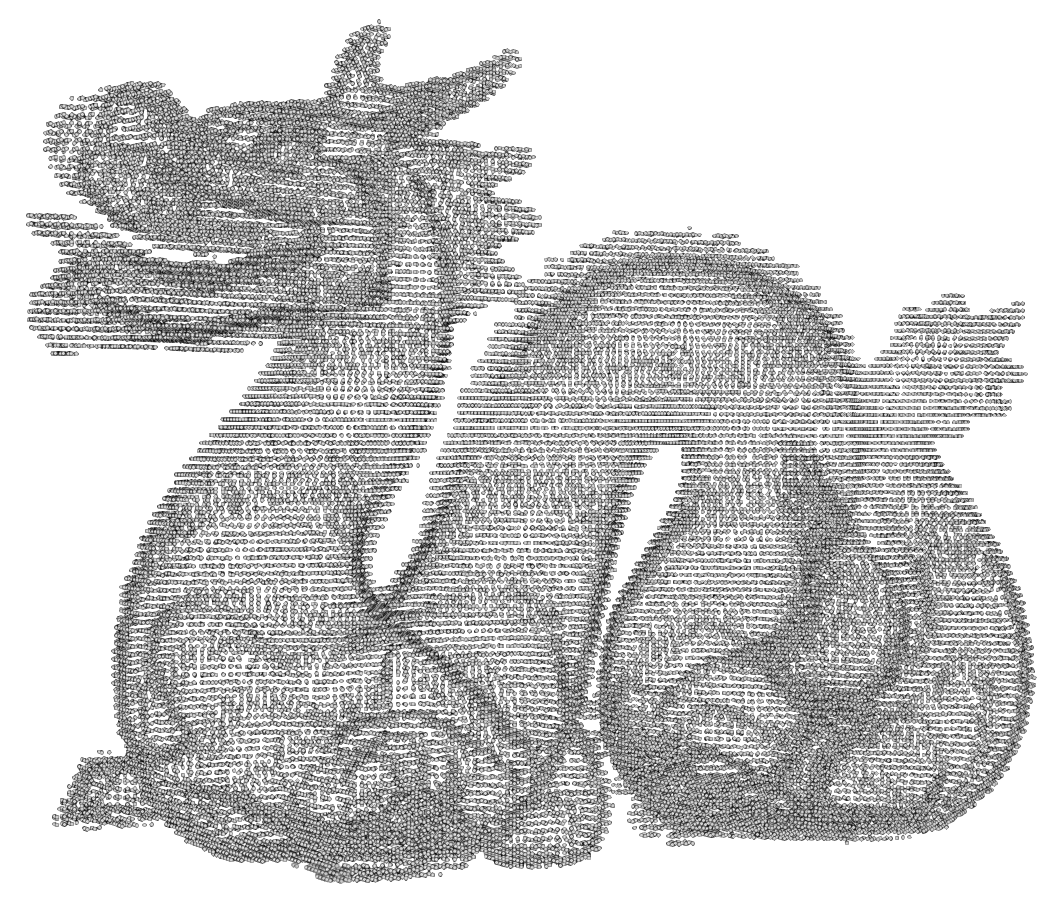}
        \subcaption{TMC13 \\ (CR=110.24, PSNR=50.5)}
        \label{fig:tmc13-dragon}
    \end{subfigure}
    \begin{subfigure}{0.3\linewidth}
        \includegraphics[width=\textwidth]{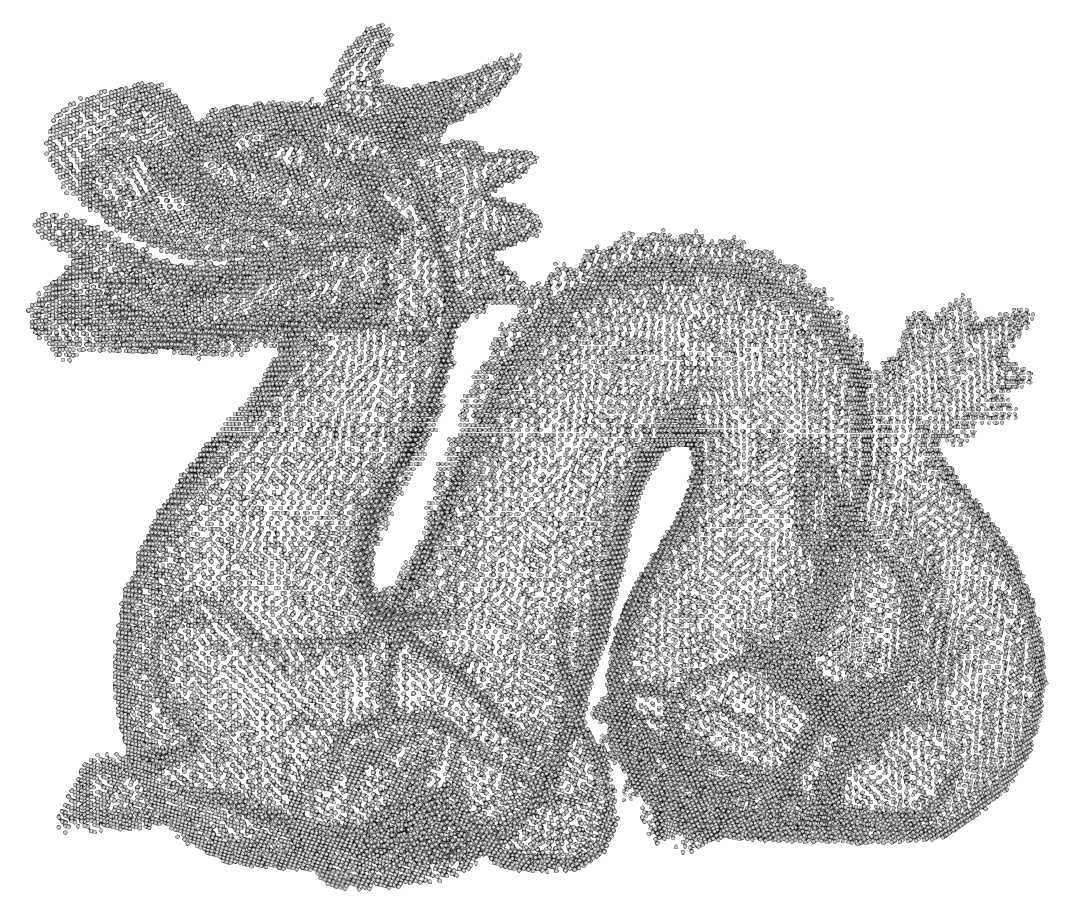}
        \subcaption{\texttt{XnYZip} \\ (CR=110.42, PSNR=52.08)}
        \label{fig:xny-dragon}
    \end{subfigure}
    \caption{
    Stanford Dragon reconstructions at a fixed CR of $110$. \texttt{XnYZip} preserves surface smoothness without the streak artefacts of LCP/Draco.
    }
    \label{fig:visual_comparison}
\end{figure}

\textbf{Visual quality.}
Beyond aggregate statistics, Figure~\ref{fig:visual_comparison} compares reconstructions of the Stanford Dragon at a fixed compression ratio of $110$. The isotropic BCC lattice yields smoother surfaces and avoids the streak-like artefacts visible in the LCP and Draco reconstructions at a comparable PSNR.

\rOne{\textbf{SFC choice.}
Table~\ref{tab:sfc-cr-comparison} compares the compression ratios of the Hilbert and Z-order variants across all datasets. The gap remains within $6.4\%$ on seven of the eight datasets; WarpX is an outlier where the highly clustered plasma coordinates make Z-order's cheaper mapping a better match for the subsequent delta+RLE stages. Overall, we recommend \texttt{XnYZip}-z as the default and \texttt{XnYZip}-h as the compression-oriented alternative for typical workloads.}


\begin{table}[t]
\centering
\footnotesize
\caption{\rOne{Compression ratio of \texttt{XnYZip}-h (Hilbert) and \texttt{XnYZip}-z (Z-order) on every dataset at $eb = 10^{-3}$, and the worst-case CR gap observed across the per-dataset $19$-eb sweep. Positive gap values indicate Hilbert wins, negative values indicate Z-order wins.}}
\label{tab:sfc-cr-comparison}
\rOne{\begin{tabular}{lrrcc}
\toprule
Dataset & CR (-h) & CR (-z) & gap @ $eb{\approx}10^{-3}$ & worst-case gap \\
\midrule
YIIP                   & 9.00   & 8.96   & $+0.4\%$  & $+2.8\%$ \\
Stanford Dragon        & 23.76  & 22.41  & $+6.0\%$  & $+6.0\%$ \\
GroEL-GroES (1AON)     & 8.55   & 8.65   & $-1.1\%$  & $+3.4\%$ \\
Vesicles               & 28.73  & 27.84  & $+3.2\%$  & $+6.4\%$ \\
USGS                   & 210.65 & 207.94 & $+1.3\%$  & $+2.1\%$ \\
EXAALT                 & 36.25  & 36.03  & $+0.6\%$  & $+2.4\%$ \\
HACC                   & 248.61 & 247.36 & $+0.5\%$  & $-2.2\%$ \\
WarpX                  & 266.52 & 277.88 & $-4.1\%$  & $-37.4\%$ \\
\bottomrule
\end{tabular}}
\end{table}


\rThree{\textbf{Database-side baselines.}
We additionally evaluate two database-oriented floating-point codecs: ALP (lossless) and BUFF (bit-precision-bounded, using the authors' \texttt{Tranway1/buff} implementation). ALP provides modest lossless gains on the four datasets where its build completes successfully, while BUFF achieves only low single-digit per-axis compression ratios within its supported bit-precision range. Both remain substantially below the lossy geometric codecs at every matched-L2 operating point in Table\ref{tab:appendix-master}, highlighting the different rate-distortion characteristics of one-dimensional column codecs and three-dimensional point-cloud compressors.}

\rTwo{\textbf{Metadata overhead at extreme CR.}
At very high CR, the bitstream becomes sufficiently small that the per-block Huffman dictionary dominates the total size: the long tail of low-frequency symbols still requires explicit representation, and the dictionary alone can account for more than $60\%$ of the bitstream, limiting the achievable CR. Replacing the metadata stream with an Exp-Golomb code (as used by TMC13 for this purpose) would avoid explicit dictionary materialization and alleviate this limitation without changing the underlying quantization geometry. We view this as an implementation-level optimization orthogonal to the main contributions of the paper.}

\subsection{Throughput}
\label{sec:throughput}

We evaluate throughput in two steps:
(1) the constant-factor overhead of TO over cube quantization, and (2) end-to-end compression and decompression throughput.

\rOne{\textbf{TO vs cube constant factor.}
We measure quantization inside the production binary with \texttt{chrono} timers and run the binary with the cube and the TO quantizer on the same input. Within the pipeline the TO call costs higher (Table~\ref{tab:to-vs-cube}) but the quantization stage is only $3$--$13\%$ of total compression time, the end-to-end overhead of TO vs cube is just $+3$--$+16\%$ wall-clock, while delivering a substantially higher CR on every dataset tested. The two-candidate parity check in TO quantization explains this constant, which is bounded and independent of $N$.}

\begin{table}[t]
\centering
\footnotesize
\caption{\rOne{Cube vs.\ TO quantization timing inside the production compressor ($eb=10^{-3}$, single-threaded). \emph{Quantize} is quantization stage time; \emph{Total} is end-to-end compression time. Each cell is the average of three runs.}}
\label{tab:to-vs-cube}
\rOne{\begin{tabular}{lrrrrr}
\toprule
                & \multicolumn{2}{c}{Quantize (ms)} & \multicolumn{2}{c}{Total (s)} & TO ratio \\
\cmidrule(lr){2-3}\cmidrule(lr){4-5}\cmidrule(lr){6-6}
Dataset         & cube & TO  & cube & TO  & TO/cube total \\
\midrule
Stanford Dragon &   3.1 &  7.9  & 0.036 & 0.038 & $1.06\times$ \\
Vesicles        &  13.7 & 38.8  & 0.305 & 0.355 & $1.16\times$ \\
USGS            & 99.8  & 340.4 & 3.39  & 3.61  & $1.07\times$ \\
HACC (22.4 M)   & 133.3 & 509.3 & 3.89  & 4.00  & $1.03\times$ \\
\bottomrule
\end{tabular}}
\end{table}

\begin{figure}[]
    \centering
    \begin{subfigure}{0.49\linewidth}
        \includegraphics[width=\textwidth]{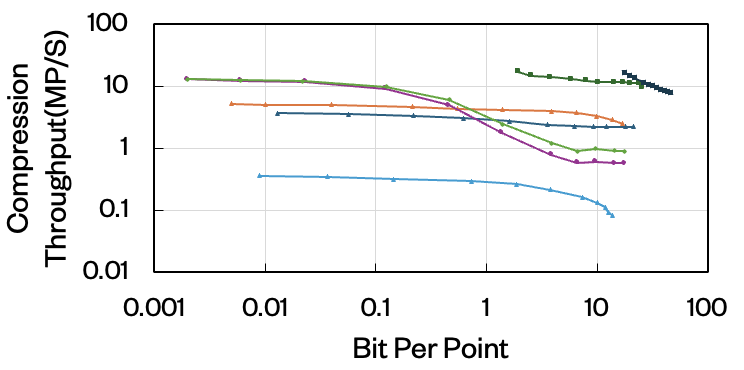}
        \subcaption{Compression Throughput on USGS}
        \vspace{1em}
        \label{fig:usgs-comp}
    \end{subfigure}
    \begin{subfigure}{0.49\linewidth}
        \includegraphics[width=\textwidth]{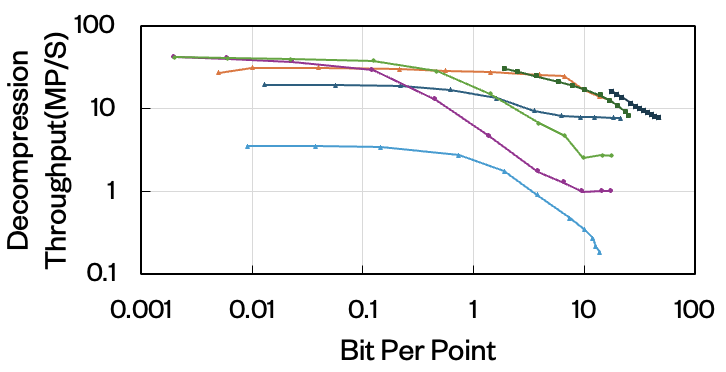}
        \subcaption{Decompression Throughput on USGS}
        \label{fig:usgs-decomp}
    \end{subfigure}
    \begin{subfigure}{0.49\linewidth}
        \includegraphics[width=\textwidth]{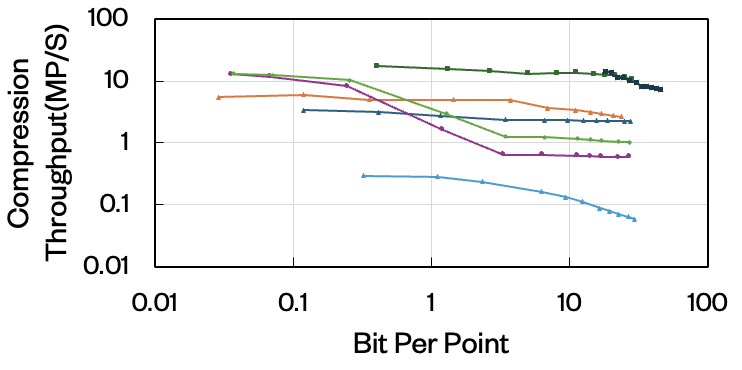}
        \subcaption{Compression Throughput on Vesicles}
        \label{fig:vesicles-comp}
    \end{subfigure}
    \begin{subfigure}{0.49\linewidth}
        \includegraphics[width=\textwidth]{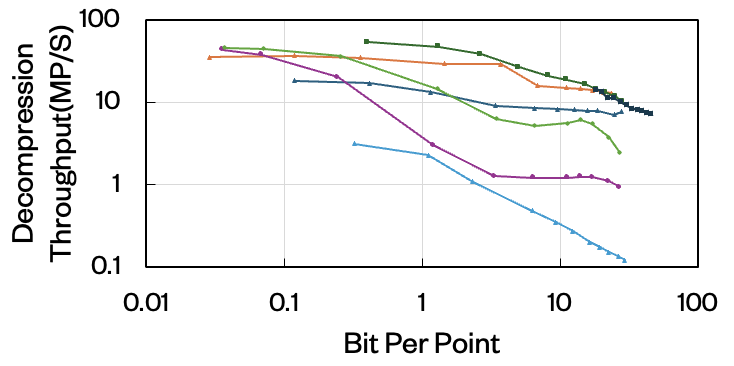}
        \subcaption{Decompression Throughput on Vesicles}
        \label{fig:vesicles-decomp}
    \end{subfigure}
    \begin{subfigure}{0.98\linewidth}
        \includegraphics[width=\textwidth]{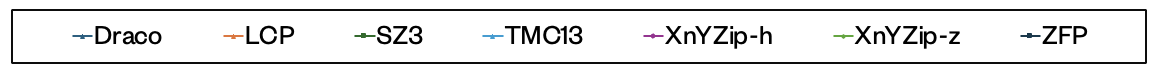}
        \label{fig:perf-legend}
    \end{subfigure}
    \caption{Throughput vs.\ bitrate on USGS and Vesicles. At low bpps (dense regime, RLE active) \texttt{XnYZip} attains the highest throughput; at higher bpps (RLE off) it remains competitive with Draco/LCP and ahead of TMC13.}
    \label{fig:throughput}
\end{figure}

\textbf{End-to-end throughput.}
\rOne{Figure~\ref{fig:throughput} plots throughput against bits-per-point on USGS and Vesicles against every baseline. The mapping and entropy coding pipeline (Huffman + Zstd) dominates the wall clock, not the quantization step.} At lower bpps, where TO quantization saturates RLE, our framework delivers the highest throughput of any dedicated point-cloud compressor evaluated; at higher bpps, where RLE is disabled, it remains competitive with Draco and LCP and consistently above TMC13.

\begin{figure}[t]
\centering
\includegraphics[width=0.95\linewidth]{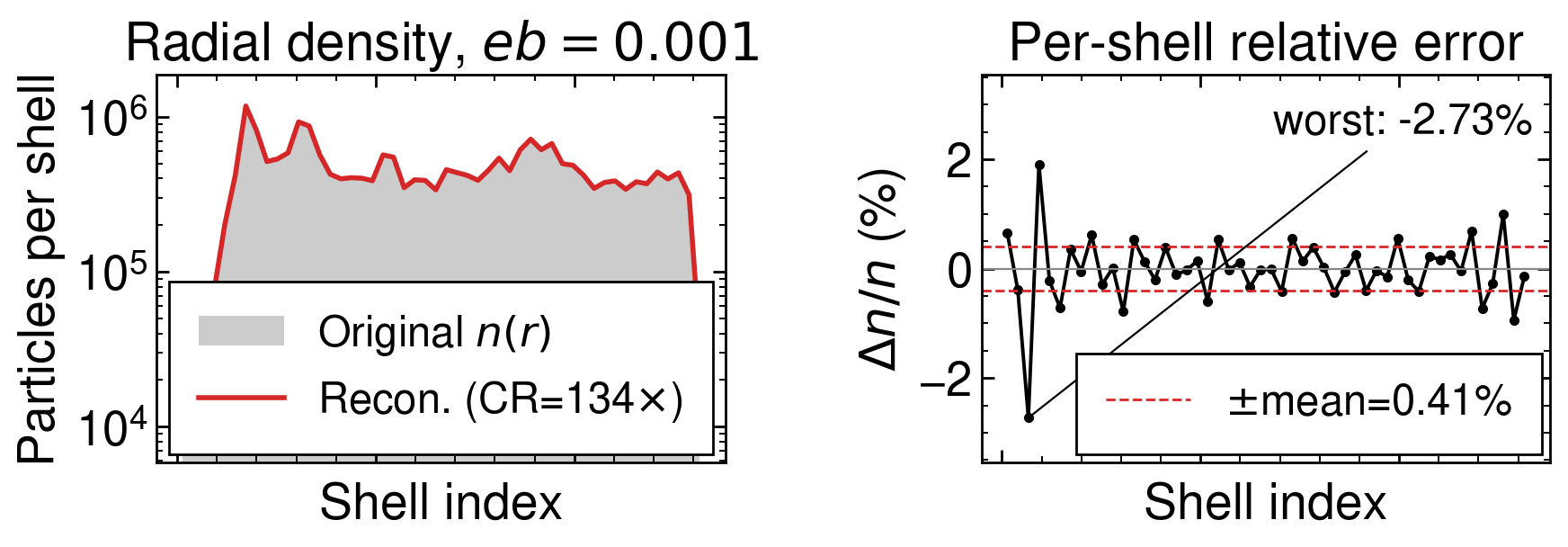}
\caption{\rThree{HACC radial density profile at $eb = 10^{-3}$ (CR $\approx 134\times$). \textbf{Left:} original particle counts per spherical shell and the reconstructed profile coincide to within line thickness across all $50$ shells. \textbf{Right:} per-shell relative error $\Delta n / n$. The mean absolute deviation is $0.41\%$; the worst single shell sits at $-2.73\%$ on a $22.4$,M-point subset.}}
\label{fig:hacc-density}
\end{figure}

\begin{table}[t]
\centering
\footnotesize
\caption{\rThree{Per-query radius-count latency on Stanford Dragon ($eb = 10^{-3}$), $50$ random ball centers. \textsc{full} = decode + expand chunks, brute-force filter. \textsc{index} = the bbox-skip + cell-count path. ``skip'' is the fraction of (chunk, query) pairs for which the sidecar bbox is disjoint from the ball; \textsc{index} counts match \textsc{full} bit-exactly on every row.}}
\label{tab:query}
\rThree{\begin{tabular}{rrrrrc}
\toprule
chunks & $r/\mathrm{diag}$ & FULL (ms) & INDEX (ms) & speedup & skip \\
\midrule
 8 & $0.02$ & $0.179$ & $0.010$ & $\mathbf{17.1\times}$ & $66\%$ \\
 8 & $0.05$ & $0.176$ & $0.017$ & $10.6\times$ & $52\%$ \\
 8 & $0.10$ & $0.181$ & $0.024$ & $7.5\times$ & $38\%$ \\
32 & $0.02$ & $0.195$ & $0.006$ & $\mathbf{34.6\times}$ & $84\%$ \\
32 & $0.05$ & $0.183$ & $0.009$ & $19.9\times$ & $74\%$ \\
32 & $0.10$ & $0.181$ & $0.016$ & $11.4\times$ & $61\%$ \\
\bottomrule
\end{tabular}}
\end{table}

\subsection{Querying the Compressed Representation}
\label{sec:case-studies}

\rThree{Compression is useful only if downstream tasks remain accurate and, ideally, operate directly on compressed data. The chunked representation in Section~\ref{sec:distributed} supports both: each chunk's RLE cell table forms a spatial histogram of cell counts, while a $28$-byte sidecar storing the bounding box and total count allows region queries to skip disjoint chunks. Radial density profiles, halo-finder occupancy maps, and radius/range filters therefore reduce to regional count queries. As a representative case, the cosmological radial density profile $n(r)$ partitions space into $N_{\text{bin}}$ concentric shells and counts particles in each shell, providing a standard input to halo, void, and overdensity analyses. We compute $n(r)$ using $50$ shells on a $22.4$,M-point HACC subset and its reconstruction at $\mathrm{CR}\approx134\times$ and $eb=10^{-3}$. The reconstructed profile differs from the original by at most $2.73\%$ per shell, with a mean deviation of $0.41\%$ (Figure~\ref{fig:hacc-density}). The largest errors occur in the innermost shells, where smaller volumes produce lower and noisier particle counts.}

\rThree{The same query runs \emph{without} full decompression: test each chunk's bbox against the query region, use the sidecar for fully contained chunks, and decode only boundary chunks. On Stanford Dragon at $eb=10^{-3}$ over $50$ random ball centers, this bit-exact path is $7.5$--$34.6\times$ faster than decode-then-filter (Table~\ref{tab:query}); the $28$-B sidecar costs $\le 0.05\%$ of the compressed bytes on realistic chunk counts and at most $1.43\%$ in a stressed $32$-chunk Dragon configuration. The same compressed bytes therefore serve as storage codec, spatial-histogram index, and per-region aggregator.}

\section{Conclusion and Future Work}
\label{sec:conclusion}

In this paper, we identify a fundamental mismatch in lossy point cloud compression: widely used L2-based distortion metrics and suboptimal cubic quantization. We formulate the problem as optimal space-filling lattice selection and show that, under MSE distortion, the BCC lattice with Truncated Octahedron Voronoi cells provides the optimal geometry. Based on this result, we develop \texttt{XnYZip}, which integrates TO-based quantization with a locality-aware SFC/RLE encoding pipeline.
Compared with state-of-the-art compressors, \texttt{XnYZip} achieves up to 3$\times$ higher compression ratios, 2.2$\times$ and 1.2$\times$ faster compression and decompression throughput, respectively, on dense scientific point cloud datasets.
Future work will explore adaptive hybrid quantization for sparse regions, multi-frame point clouds, and parallel CPU/GPU implementations for high-throughput \textit{in-situ} compression.
\section{Acknowledgement}
The material was supported by the U.S. Department of Energy, Office of Science, Advanced Scientific Computing Research (ASCR), under contract DE-AC02-06CH11357, and supported by the U.S. National Science Foundation under Grant OAC-2311875, OAC-2344717, OAC-2513768, OAC-2514036, OAC-2514034, OAC-2609480, OAC-2542646, and III-2107213. We acknowledge the computing resources provided on Bebop (operated by Laboratory Computing Resource Center at Argonne).

\newpage
\balance
\bibliographystyle{ACM-Reference-Format}
\bibliography{refs}

@article{alp,
author = {Afroozeh, Azim and Kuffo, Leonardo X. and Boncz, Peter},
title = {ALP: Adaptive Lossless floating-Point Compression},
year = {2023},
issue_date = {December 2023},
publisher = {Association for Computing Machinery},
address = {New York, NY, USA},
volume = {1},
number = {4},
url = {https://doi.org/10.1145/3626717},
doi = {10.1145/3626717},
journal = {Proc. ACM Manag. Data},
month = dec,
articleno = {230},
numpages = {26}
}

@INPROCEEDINGS {tspsz,
author = { Xia, Mingze and Wang, Bei and Li, Yuxiao and Jiao, Pu and Liang, Xin and Guo, Hanqi },
booktitle = { 2025 IEEE 41st International Conference on Data Engineering (ICDE) },
title = {{ TspSZ: An Efficient Parallel Error-Bounded Lossy Compressor for Topological Skeleton Preservation }},
year = {2025},
volume = {},
ISSN = {},
pages = {3682-3695},
doi = {10.1109/ICDE65448.2025.00275},
url = {https://doi.ieeecomputersociety.org/10.1109/ICDE65448.2025.00275},
publisher = {IEEE Computer Society},
address = {Los Alamitos, CA, USA},
month =May}

@article{buff,
author = {Liu, Chunwei and Jiang, Hao and Paparrizos, John and Elmore, Aaron J.},
title = {Decomposed bounded floats for fast compression and queries},
year = {2021},
issue_date = {July 2021},
publisher = {VLDB Endowment},
volume = {14},
number = {11},
issn = {2150-8097},
url = {https://doi.org/10.14778/3476249.3476305},
doi = {10.14778/3476249.3476305},
journal = {Proc. VLDB Endow.},
month = jul,
pages = {2586–2598},
numpages = {13}
}

@article{fpsurvey,
author = {Hishida, Kaisei and Liu, Chunwei and Paparrizos, John and Elmore, Aaron J.},
title = {Beyond Compression: A Comprehensive Evaluation of Lossless Floating-Point Compression},
year = {2025},
issue_date = {July 2025},
publisher = {VLDB Endowment},
volume = {18},
number = {11},
issn = {2150-8097},
url = {https://doi.org/10.14778/3749646.3749701},
doi = {10.14778/3749646.3749701},
journal = {Proc. VLDB Endow.},
month = jul,
pages = {4396–4409},
numpages = {14}
}

@misc{xia2026timevaryingvectorfieldcompression,
      title={Time-varying Vector Field Compression with Preserved Critical Point Trajectories}, 
      author={Mingze Xia and Yuxiao Li and Pu Jiao and Bei Wang and Xin Liang and Hanqi Guo},
      year={2026},
      eprint={2510.25143},
      archivePrefix={arXiv},
      primaryClass={cs.DB},
      url={https://arxiv.org/abs/2510.25143}, 
}

@INPROCEEDINGS{hu2025dcc,
  author={Chen, Yufan and Zou, Xiangyu and Deng, Kaiwen and Hu, Hao and Deng, Cai and Feng, Ke and Xia, Wen},
  booktitle={2025 Data Compression Conference (DCC)}, 
  title={Apic: A Precomputation-Based Integer Compressor for OLTP Databases}, 
  year={2025},
  volume={},
  number={},
  pages={303-312},
  doi={10.1109/DCC62719.2025.00038}}

@INPROCEEDINGS{hu2025icde,
  author={Hu, Hao and Zheng, Qiyang and Zou, Xiangyu and Qin, Lisha and Zhang, Chengwei and Zhang, Wanchuan and Jiang, Zhaoheng and Tao, Dingwen and Wang, Hongpeng and Xia, Wen},
  booktitle={2025 IEEE 41st International Conference on Data Engineering (ICDE)}, 
  title={A Cost-Effective and Decompression-Transparent Compressor for OLTP-Oriented Databases}, 
  year={2025},
  volume={},
  number={},
  pages={405-418},
  doi={10.1109/ICDE65448.2025.00037}}

@INPROCEEDINGS{xia2024,
  author={Xia, Mingze and Di, Sheng and Cappello, Franck and Jiao, Pu and Zhao, Kai and Liu, Jinyang and Wu, Xuan and Liang, Xin and Guo, Hanqi},
  booktitle={2024 IEEE 40th International Conference on Data Engineering (ICDE)}, 
  title={Preserving Topological Feature with Sign-of-Determinant Predicates in Lossy Compression: A Case Study of Vector Field Critical Points}, 
  year={2024},
  volume={},
  number={},
  pages={4979-4992},
  doi={10.1109/ICDE60146.2024.00378}}

@inproceedings{summarystore,
author = {Agrawal, Nitin and Vulimiri, Ashish},
title = {Low-Latency Analytics on Colossal Data Streams with SummaryStore},
year = {2017},
isbn = {9781450350853},
publisher = {Association for Computing Machinery},
address = {New York, NY, USA},
url = {https://doi.org/10.1145/3132747.3132758},
doi = {10.1145/3132747.3132758},
booktitle = {Proceedings of the 26th Symposium on Operating Systems Principles},
pages = {647–664},
numpages = {18},
location = {Shanghai, China},
series = {SOSP '17}
}

@article{qpet,
author = {Liu, Jinyang and Jiao, Pu and Zhao, Kai and Liang, Xin and Di, Sheng and Cappello, Franck},
title = {QPET: A Versatile and Portable Quantity-of-Interest-Preservation Framework for Error-Bounded Lossy Compression},
year = {2025},
issue_date = {April 2025},
publisher = {VLDB Endowment},
volume = {18},
number = {8},
issn = {2150-8097},
url = {https://doi.org/10.14778/3742728.3742739},
doi = {10.14778/3742728.3742739},
journal = {Proc. VLDB Endow.},
month = apr,
pages = {2440–2453},
numpages = {14}
}

@INPROCEEDINGS{jin2022improving,
  author={Jin, Sian and Di, Sheng and Tian, Jiannan and Byna, Suren and Tao, Dingwen and Cappello, Franck},
  booktitle={2022 IEEE 38th International Conference on Data Engineering (ICDE)}, 
  title={Improving Prediction-Based Lossy Compression Dramatically via Ratio-Quality Modeling}, 
  year={2022},
  volume={},
  number={},
  pages={2494-2507},
  doi={10.1109/ICDE53745.2022.00232}}

@ARTICLE{Gersho1979,
  author={Gersho, A.},
  journal={IEEE Transactions on Information Theory}, 
  title={Asymptotically optimal block quantization}, 
  year={1979},
  volume={25},
  number={4},
  pages={373-380},
  doi={10.1109/TIT.1979.1056067}}

@ARTICLE{VPCC,
  author={Schwarz, Sebastian and Preda, Marius and Baroncini, Vittorio and Budagavi, Madhukar and Cesar, Pablo and Chou, Philip A. and Cohen, Robert A. and Krivokuća, Maja and Lasserre, Sébastien and Li, Zhu and Llach, Joan and Mammou, Khaled and Mekuria, Rufael and Nakagami, Ohji and Siahaan, Ernestasia and Tabatabai, Ali and Tourapis, Alexis M. and Zakharchenko, Vladyslav},
  journal={IEEE Journal on Emerging and Selected Topics in Circuits and Systems}, 
  title={Emerging MPEG Standards for Point Cloud Compression}, 
  year={2019},
  volume={9},
  number={1},
  pages={133-148},
  doi={10.1109/JETCAS.2018.2885981}}

@inproceedings{Huang2006,
booktitle = {Symposium on Point-Based Graphics},
editor = {Mario Botsch and Baoquan Chen and Mark Pauly and Matthias Zwicker},
title = {{Octree-Based Progressive Geometry Coding of Point Clouds}},
author = {Huang, Yan and Peng, Jingliang and Kuo, C.-C. Jay and Gopi, M.},
year = {2006},
publisher = {The Eurographics Association},
ISSN = {1811-7813},
ISBN = {3-905673-32-0},
DOI = {/10.2312/SPBG/SPBG06/103-110}
}

@INPROCEEDINGS{Kathariya2018,
  author={Kathariya, Birendra and Li, Li and Li, Zhu and Alvarez, Jose and Chen, Jianle},
  booktitle={2018 IEEE International Conference on Multimedia and Expo (ICME)}, 
  title={Scalable Point Cloud Geometry Coding with Binary Tree Embedded Quadtree}, 
  year={2018},
  volume={},
  number={},
  pages={1-6},
  doi={10.1109/ICME.2018.8486481}}

@inproceedings{Kitago2006,
booktitle = {Symposium on Point-Based Graphics},
editor = {Mario Botsch and Baoquan Chen and Mark Pauly and Matthias Zwicker},
title = {{Efficient and Prioritized Point Subsampling for CSRBF Compression}},
author = {Kitago, Masaki and Gopi, M.},
year = {2006},
publisher = {The Eurographics Association},
ISSN = {1811-7813},
ISBN = {3-905673-32-0},
DOI = {/10.2312/SPBG/SPBG06/121-128}
}

@INPROCEEDINGS{Kammerl2012,
  author={Kammerl, Julius and Blodow, Nico and Rusu, Radu Bogdan and Gedikli, Suat and Beetz, Michael and Steinbach, Eckehard},
  booktitle={2012 IEEE International Conference on Robotics and Automation}, 
  title={Real-time compression of point cloud streams}, 
  year={2012},
  volume={},
  number={},
  pages={778-785},
  doi={10.1109/ICRA.2012.6224647}}

@article{Taubin1998,
author = {Taubin, Gabriel and Rossignac, Jarek},
title = {Geometric compression through topological surgery},
year = {1998},
issue_date = {April 1998},
publisher = {Association for Computing Machinery},
address = {New York, NY, USA},
volume = {17},
number = {2},
issn = {0730-0301},
url = {https://doi.org/10.1145/274363.274365},
doi = {10.1145/274363.274365},
journal = {ACM Trans. Graph.},
month = apr,
pages = {84–115},
numpages = {32}
}

@inproceedings{Ochotta2004,
author = {Ochotta, Tilo and Saupe, Dietmar},
title = {Compression of point-based 3D models by shape-adaptive wavelet coding of multi-height fields},
year = {2004},
isbn = {3905673096},
publisher = {Eurographics Association},
address = {Goslar, DEU},
booktitle = {Proceedings of the First Eurographics Conference on Point-Based Graphics},
pages = {103–112},
numpages = {10},
location = {Switzerland},
series = {SPBG'04}
}

@Article{Lien2010,
author={Lien, Jyh-Ming
and Kurillo, Gregorij
and Bajcsy, Ruzena},
title={Multi-camera tele-immersion system with real-time model driven data compression},
journal={The Visual Computer},
year={2010},
month={Jan},
day={01},
volume={26},
number={1},
pages={3-15},
issn={1432-2315},
doi={10.1007/s00371-009-0367-8},
url={https://doi.org/10.1007/s00371-009-0367-8}
}

@Article{Daribo2012,
author={Daribo, Ismael
and Furukawa, Ryo
and Sagawa, Ryusuke
and Kawasaki, Hiroshi
and Hiura, Shinsaku
and Asada, Naoki},
title={Efficient rate-distortion compression of dynamic point cloud for grid-pattern-based 3D scanning systems},
journal={3D Research},
year={2012},
month={Jan},
day={17},
volume={3},
number={1},
pages={2},
issn={2092-6731},
doi={10.1007/3DRes.01(2012)2},
url={https://doi.org/10.1007/3DRes.01(2012)2}
}

@INPROCEEDINGS{Merkle2007,
  author={Merkle, Philipp and Smolic, Aljoscha and Muller, Karsten and Wiegand, Thomas},
  booktitle={2007 IEEE International Conference on Image Processing}, 
  title={Multi-View Video Plus Depth Representation and Coding}, 
  year={2007},
  volume={1},
  number={},
  pages={I - 201-I - 204},
  doi={10.1109/ICIP.2007.4378926}}

@inproceedings{Gumhold2005,
author = {Gumhold, Stefan and Kami, Zachi and Isenburg, Martin and Seidel, Hans-Peter},
title = {Predictive point-cloud compression},
year = {2005},
isbn = {9781450378277},
publisher = {Association for Computing Machinery},
address = {New York, NY, USA},
url = {https://doi.org/10.1145/1187112.1187277},
doi = {10.1145/1187112.1187277},
booktitle = {ACM SIGGRAPH 2005 Sketches},
pages = {137–es},
location = {Los Angeles, California},
series = {SIGGRAPH '05}
}

@INPROCEEDINGS{2017fan,
  author={Fan, Haoqiang and Su, Hao and Guibas, Leonidas},
  booktitle={2017 IEEE Conference on Computer Vision and Pattern Recognition (CVPR)}, 
  title={A Point Set Generation Network for 3D Object Reconstruction from a Single Image}, 
  year={2017},
  volume={},
  number={},
  pages={2463-2471},
  doi={10.1109/CVPR.2017.264}}

@misc{MPEG_I_PCC,
  author       = {MPEG},
  title        = {{MPEG-I Part 9: Geometry-based Point Cloud Compression}},
  howpublished = {\url{https://www.mpeg.org/standards/MPEG-I/9/}},
  note         = {Accessed: 2024-03-22},
  year         = {2024}
}

@inproceedings{cao2019survey,
author = {Cao, Chao and Preda, Marius and Zaharia, Titus},
title = {3D Point Cloud Compression: A Survey},
year = {2019},
isbn = {9781450367981},
publisher = {Association for Computing Machinery},
address = {New York, NY, USA},
url = {https://doi.org/10.1145/3329714.3338130},
doi = {10.1145/3329714.3338130},
booktitle = {Proceedings of the 24th International Conference on 3D Web Technology},
pages = {1–9},
numpages = {9},
location = {LA, CA, USA},
series = {Web3D '19}
}

@article{LCP,
author = {Zhang, Longtao and Li, Ruoyu and Ren, Congrong and Di, Sheng and Liu, Jinyang and Huang, Jiajun and Underwood, Robert and Grosset, Pascal and Tao, Dingwen and Liang, Xin and Guo, Hanqi and Cappello, Franck and Zhao, Kai},
title = {LCP: Enhancing Scientific Data Management with Lossy Compression for Particles},
year = {2025},
issue_date = {February 2025},
publisher = {Association for Computing Machinery},
address = {New York, NY, USA},
volume = {3},
number = {1},
url = {https://doi.org/10.1145/3709700},
doi = {10.1145/3709700},
journal = {Proc. ACM Manag. Data},
month = feb,
articleno = {50},
numpages = {27}
}

@ARTICLE{voronoi,
  author={Conway, J. and Sloane, N.},
  journal={IEEE Transactions on Information Theory}, 
  title={Voronoi regions of lattices, second moments of polytopes, and quantization}, 
  year={1982},
  volume={28},
  number={2},
  pages={211-226},
  doi={10.1109/TIT.1982.1056483}}

@misc{EXAALT2021,
  author       = {{EXAALT Project}},
  title        = {{EXAALT: Molecular Dynamics at Exascale for Materials Science}},
  year         = {2021},
  howpublished = {\url{https://www.exascaleproject.org/research-project/exaalt/}},
  note         = {Online}
}

@misc{YIIP,
  author       = {Irfan, Alibay and Oliver, Beckstein and Shujie, Fan and Richard, J. Gowers and Micaela, Matta and Lily, Wang},
  title        = {YiiP equilibrium dataset},
  year         = {2018},
  howpublished = {\url{https://www.mdanalysis.org/MDAnalysisData/yiip_equilibrium.html}},
  note         = {Online}
}

@article{rabitq,
author = {Gao, Jianyang and Long, Cheng},
title = {RaBitQ: Quantizing High-Dimensional Vectors with a Theoretical Error Bound for Approximate Nearest Neighbor Search},
year = {2024},
issue_date = {June 2024},
publisher = {Association for Computing Machinery},
address = {New York, NY, USA},
volume = {2},
number = {3},
url = {https://doi.org/10.1145/3654970},
doi = {10.1145/3654970},
journal = {Proc. ACM Manag. Data},
month = may,
articleno = {167},
numpages = {27}
}

@inproceedings{oracle,
author = {Poess, Meikel and Potapov, Dmitry},
title = {Data compression in Oracle},
year = {2003},
isbn = {0127224424},
publisher = {VLDB Endowment},
booktitle = {Proceedings of the 29th International Conference on Very Large Data Bases - Volume 29},
pages = {937–947},
numpages = {11},
location = {Berlin, Germany},
series = {VLDB '03}
}

@inproceedings{compInDB,
author = {Iyer, Balakrishna R. and Wilhite, David},
title = {Data Compression Support in Databases},
year = {1994},
isbn = {1558601538},
publisher = {Morgan Kaufmann Publishers Inc.},
address = {San Francisco, CA, USA},
booktitle = {Proceedings of the 20th International Conference on Very Large Data Bases},
pages = {695–704},
numpages = {10},
series = {VLDB '94}
}

@article{Jiao2022,
author = {Jiao, Pu and Di, Sheng and Guo, Hanqi and Zhao, Kai and Tian, Jiannan and Tao, Dingwen and Liang, Xin and Cappello, Franck},
title = {Toward Quantity-of-Interest Preserving Lossy Compression for Scientific Data},
year = {2022},
issue_date = {December 2022},
publisher = {VLDB Endowment},
volume = {16},
number = {4},
issn = {2150-8097},
url = {https://doi.org/10.14778/3574245.3574255},
doi = {10.14778/3574245.3574255},
journal = {Proc. VLDB Endow.},
month = dec,
pages = {697–710},
numpages = {14}
}

@article{Tropf1981MultimensionalRS,
  title={Multimensional Range Search in Dynamically Balanced Trees},
  author={Hermann Tropf and H. Herzog},
  journal={Angew. Inform.},
  year={1981},
  volume={23},
  pages={71-77},
  url={https://api.semanticscholar.org/CorpusID:26857103}
}

@techreport{morton1966,
  author       = {Morton, G. M.},
  title        = {A Computer Oriented Geodetic Data Base; and a New Technique in File Sequencing},
  year         = {1966},
  institution  = {IBM Ltd.},
  address      = {Ottawa, Canada},
  note         = {Technical Report}
}

@inproceedings{Michael1995,
author = {Deering, Michael},
title = {Geometry compression},
year = {1995},
isbn = {0897917014},
publisher = {Association for Computing Machinery},
address = {New York, NY, USA},
url = {https://doi.org/10.1145/218380.218391},
doi = {10.1145/218380.218391},
booktitle = {Proceedings of the 22nd Annual Conference on Computer Graphics and Interactive Techniques},
pages = {13–20},
numpages = {8},
series = {SIGGRAPH '95}
}

@misc{jpeg1992short,
  author       = {{Joint Photographic Experts Group}},
  title        = {JPEG: Still Image Data Compression Standard (ISO/IEC 10918-1)},
  year         = {1992},
  howpublished = {ITU-T Recommendation T.81 and ISO/IEC 10918-1 Standard},
  note         = {Online; accessed on 2025-10-17}
}

@misc{usgs3dep2024,
  author       = {{U.S. Geological Survey}},
  title        = {USGS 3DEP LiDAR Point Cloud Dataset},
  howpublished = {\url{https://www.usgs.gov/news/technical-announcement/usgs-3dep-lidar-point-cloud-now-available-amazon-public-dataset/}},
  year         = {2024},
  note         = {Online; }
}

@article{Kenney2015,
author = "Ian M. Kenney and Oliver Beckstein",
title = "{Technical Report: SPIDAL Summer REU 2015: Biomolecular benchmark systems}",
year = "2015",
month = "10",
url = "https://figshare.com/articles/journal_contribution/Technical_Report_SPIDAL_Summer_REU_2015_Biomolecular_benchmark_systems/1588804",
doi = "10.6084/m9.figshare.1588804.v1"
}

@Article{Xu1997,
author={Xu, Zhaohui
and Horwich, Arthur L.
and Sigler, Paul B.},
title={The crystal structure of the asymmetric GroEL--GroES--(ADP)7 chaperonin complex},
journal={Nature},
year={1997},
month={Aug},
day={01},
volume={388},
number={6644},
pages={741-750},
issn={1476-4687},
doi={10.1038/41944},
url={https://doi.org/10.1038/41944}
}

@misc{stanford3dscan,
  author       = {{Stanford University Computer Graphics Laboratory}},
  title        = {The Stanford 3D Scanning Repository},
  howpublished = {\url{https://graphics.stanford.edu/data/3Dscanrep/}},
  year         = {1996},
  note         = {Online; }
}

@book{Cover2006,
author = {Cover, Thomas M. and Thomas, Joy A.},
title = {Elements of Information Theory (Wiley Series in Telecommunications and Signal Processing)},
year = {2006},
isbn = {0471241954},
publisher = {Wiley-Interscience},
address = {USA}
}

@Article{Hilbert1891,
author={Hilbert, David},
title={Ueber die stetige Abbildung einer Line auf ein Fl{\"a}chenst{\"u}ck},
journal={Mathematische Annalen},
year={1891},
month={Sep},
day={01},
volume={38},
number={3},
pages={459-460},
issn={1432-1807},
doi={10.1007/BF01199431},
url={https://doi.org/10.1007/BF01199431}
}

@ARTICLE{Schwarz2019GPCC,
  author={Schwarz, Sebastian and Preda, Marius and Baroncini, Vittorio and Budagavi, Madhukar and Cesar, Pablo and Chou, Philip A. and Cohen, Robert A. and Krivokuća, Maja and Lasserre, Sébastien and Li, Zhu and Llach, Joan and Mammou, Khaled and Mekuria, Rufael and Nakagami, Ohji and Siahaan, Ernestasia and Tabatabai, Ali and Tourapis, Alexis M. and Zakharchenko, Vladyslav},
  journal={IEEE Journal on Emerging and Selected Topics in Circuits and Systems}, 
  title={Emerging MPEG Standards for Point Cloud Compression}, 
  year={2019},
  volume={9},
  number={1},
  pages={133-148},
  doi={10.1109/JETCAS.2018.2885981}}

@article{Kersten2011,
author = {Kersten, Martin L. and Idreos, Stratos and Manegold, Stefan and Liarou, Erietta},
title = {The researcher's guide to the data deluge: querying a scientific database in just a few seconds},
year = {2011},
issue_date = {August 2011},
publisher = {VLDB Endowment},
volume = {4},
number = {12},
issn = {2150-8097},
url = {https://doi.org/10.14778/3402755.3402799},
doi = {10.14778/3402755.3402799},
journal = {Proc. VLDB Endow.},
month = aug,
pages = {1474–1477},
numpages = {4}
}

@article{Gray2005,
author = {Gray, Jim and Liu, David T. and Nieto-Santisteban, Maria and Szalay, Alex and DeWitt, David J. and Heber, Gerd},
title = {Scientific data management in the coming decade},
year = {2005},
issue_date = {December 2005},
publisher = {Association for Computing Machinery},
address = {New York, NY, USA},
volume = {34},
number = {4},
issn = {0163-5808},
url = {https://doi.org/10.1145/1107499.1107503},
doi = {10.1145/1107499.1107503},
journal = {SIGMOD Rec.},
month = dec,
pages = {34–41},
numpages = {8}
}

@inproceedings{Cheng2014,
author = {Cheng, Yu and Rusu, Florin},
title = {Parallel in-situ data processing with speculative loading},
year = {2014},
isbn = {9781450323765},
publisher = {Association for Computing Machinery},
address = {New York, NY, USA},
url = {https://doi.org/10.1145/2588555.2593673},
doi = {10.1145/2588555.2593673},
booktitle = {Proceedings of the 2014 ACM SIGMOD International Conference on Management of Data},
pages = {1287–1298},
numpages = {12},
location = {Snowbird, Utah, USA},
series = {SIGMOD '14}
}

@ARTICLE{hires,
  author={Lookabaugh, T.D. and Gray, R.M.},
  journal={IEEE Transactions on Information Theory}, 
  title={High-resolution quantization theory and the vector quantizer advantage}, 
  year={1989},
  volume={35},
  number={5},
  pages={1020-1033},
  doi={10.1109/18.42217}}

@misc{google_draco,
  author = {Google},
  title = {Draco: A library for compressing and decompressing 3D geometric meshes and point clouds.},
  year = {2017},
  howpublished = {\url{https://github.com/google/draco}},
  note = {Online}

}

@misc{mpeg_pcc_tmc13,
  author = {{MPEG Group}},
  title = {Geometry based point cloud compression (G-PCC) test model},
  year = {2017},
  howpublished = {\url{https://github.com/MPEGGroup/mpeg-pcc-tmc13}},
  note = {Online}
}

@misc{cesm-atm,
author={{Community Earth System Model (CESM) Atmosphere Model}},
howpublished = {\url{http://www.cesm.ucar.edu/models/}},
note={Online},
year={2019}
}

@misc{nyx,
author={{NYX simulation}},
howpublished = {\url{https://amrex-astro.github.io/Nyx/}},
note={Online},
year={2019}
}

@INPROCEEDINGS {warpx,
author = {L. Fedeli and A. Huebl and F. Boillod-Cerneux and T. Clark and K. Gott and C. Hillairet and S. Jaure and A. Leblanc and R. Lehe and A. Myers and C. Piechurski and M. Sato and N. Zaim and W. Zhang and J. Vay and H. Vincenti},
booktitle = {SC22: International Conference for High Performance Computing, Networking, Storage and Analysis},
title = {Pushing the Frontier in the Design of Laser-Based Electron Accelerators with Groundbreaking Mesh-Refined Particle-In-Cell Simulations on Exascale-Class Supercomputers},
year = {2022},
volume = {},
issn = {},
pages = {1-12},
doi = {10.1109/SC41404.2022.00008},
url = {https://doi.ieeecomputersociety.org/10.1109/SC41404.2022.00008},
publisher = {IEEE Computer Society},
address = {Los Alamitos, CA, USA},
month = {nov}
}

@misc{HACC,
author={{HACC team (ECP EXASKY)}},
howpublished = {\url{https://sdrbench.github.io/}},
note={Online},
year={2019}
}

@article{modelarDB,
author = {Jensen, S\o{}ren Kejser and Pedersen, Torben Bach and Thomsen, Christian},
title = {ModelarDB: modular model-based time series management with spark and cassandra},
year = {2018},
issue_date = {July 2018},
publisher = {VLDB Endowment},
volume = {11},
number = {11},
issn = {2150-8097},
url = {https://doi.org/10.14778/3236187.3236215},
doi = {10.14778/3236187.3236215},
journal = {Proc. VLDB Endow.},
month = jul,
pages = {1688–1701},
numpages = {14}
}

@article{wallace1992jpeg,
  title={The {JPEG} still picture compression standard},
  author={Wallace, Gregory K},
  journal={IEEE Transactions on Consumer Electronics},
  volume={38},
  number={1},
  pages={xviii--xxxiv},
  year={1992},
  publisher={IEEE}
}

@article{zfp,
  title={Fixed-rate compressed floating-point arrays},
  author={Lindstrom, Peter},
  journal={IEEE Transactions on Visualization and Computer Graphics},
  volume={20},
  number={12},
  pages={2674--2683},
  year={2014},
  publisher={IEEE}
}

@book{taubman2012jpeg2000,
  title={{JPEG2000} image compression fundamentals, standards and practice: image compression fundamentals, standards and practice},
  author={Taubman, David and Marcellin, Michael},
  volume={642},
  year={2012},
  publisher={Springer Science \& Business Media}
}

@misc{cuZFP,
  author={{cuZFP}},
  howpublished = {\url{https://github.com/LLNL/zfp/tree/develop/src/cuda_zfp}},
  year = {2023},
  note = {Online}
}

@inproceedings{sz16,
  title={Fast error-bounded lossy HPC data compression with {SZ}},
  author={Di, Sheng and Cappello, Franck},
  booktitle={2016 IEEE International Parallel and Distributed Processing Symposium},
  pages={730--739},
  year={2016},
  organization={IEEE},
  publisher={IEEE},
  address={Chicago, IL, USA}
}

@inproceedings{wang2022tac,
  title={TAC: Optimizing Error-Bounded Lossy Compression for Three-Dimensional Adaptive Mesh Refinement Simulations},
  author={Wang, Daoce and Pulido, Jesus and Grosset, Pascal and Jin, Sian and Tian, Jiannan and Ahrens, James and Tao, Dingwen},
  booktitle={Proceedings of the 31st International Symposium on High-Performance Parallel and Distributed Computing},
  pages={135--147},
  year={2022}
}

@inproceedings{sz18,
  title={Error-controlled lossy compression optimized for high compression ratios of scientific datasets},
  author={Liang, Xin and Di, Sheng and Tao, Dingwen and Li, Sihuan and Li, Shaomeng and Guo, Hanqi and Chen, Zizhong and Cappello, Franck},
  booktitle={2018 IEEE International Conference on Big Data},
  pages={438--447},
  year={2018},
  organization={IEEE}
}

@online{hdf5filter-sz,
  author={Di, Sheng},
  title={{H5Z-SZ}},
  url = {https://github.com/disheng222/H5Z-SZ},
  year = {2023},
  note = {Online}
}

@online{hdf5-filter,
  title={HDF5 Filters},
  url = {https://docs.hdfgroup.org/hdf5/develop/_f_i_l_t_e_r.html},
  year = {2023},
  note = {Online}
}

@ONLINE{hdf5,
    author = {{The HDF Group}},
    title = {Hierarchical data format version 5},
    url = {http://www.hdfgroup.org/HDF5},
    year = {2023},
    note = {Online}
}

@ARTICLE{sz3,
  author={Liang, Xin and Zhao, Kai and Di, Sheng and Li, Sihuan and Underwood, Robert and Gok, Ali M. and Tian, Jiannan and Deng, Junjing and Calhoun, Jon C. and Tao, Dingwen and Chen, Zizhong and Cappello, Franck},
  journal={IEEE Transactions on Big Data}, 
  title={SZ3: A Modular Framework for Composing Prediction-Based Error-Bounded Lossy Compressors}, 
  year={2022},
  volume={},
  number={},
  pages={1-14},
  doi={10.1109/TBDATA.2022.3201176}
}

@inproceedings{sz17,
  title={Significantly improving lossy compression for scientific data sets based on multidimensional prediction and error-controlled quantization},
  author={Tao, Dingwen and Di, Sheng and Chen, Zizhong and Cappello, Franck},
  booktitle={2017 IEEE International Parallel and Distributed Processing Symposium},
  pages={1129--1139},
  year={2017},
  organization={IEEE}
}

@article{ainsworth2018multilevel,
  title={Multilevel techniques for compression and reduction of scientific data—the univariate case},
  author={Ainsworth, Mark and Tugluk, Ozan and Whitney, Ben and Klasky, Scott},
  journal={Computing and Visualization in Science},
  volume={19},
  number={5--6},
  pages={65--76},
  year={2018},
  publisher={Springer}
}

@inproceedings{ZhiyuanGK01,
author = {Chen, Zhiyuan and Gehrke, Johannes and Korn, Flip},
title = {Query optimization in compressed database systems},
year = {2001},
isbn = {1581133324},
pages = {271–282},
numpages = {12}
}

@INPROCEEDINGS{Fenget21,
  author={Zhang, Feng and Pan, Zaifeng and Zhou, Yanliang and Zhai, Jidong and Shen, Xipeng and Mutlu, Onur and Du, Xiaoyong},
  booktitle={ICDE}, 
  title={G-TADOC: Enabling Efficient GPU-Based Text Analytics without Decompression}, 
  year={2021},
  volume={},
  number={},
  pages={1679-1690}
}

@article{Fenget18,
author = {Zhang, Feng and Zhai, Jidong and Shen, Xipeng and Mutlu, Onur and Chen, Wenguang},
title = {Efficient document analytics on compressed data: method, challenges, algorithms, insights},
year = {2018},
issue_date = {July 2018},
volume = {11},
number = {11},
issn = {2150-8097},
journal = {Proc. VLDB Endow.},
month = jul,
pages = {1522–1535},
numpages = {14}
}

@ARTICLE{jegou2011pq,
  author={Jégou, Herve and Douze, Matthijs and Schmid, Cordelia},
  journal={IEEE Transactions on Pattern Analysis and Machine Intelligence}, 
  title={Product Quantization for Nearest Neighbor Search}, 
  year={2011},
  volume={33},
  number={1},
  pages={117-128}
}


\end{document}